\documentclass[preprint]{aastex} 
\usepackage{graphicx}
\usepackage{epsf}
\usepackage{enumerate}
\usepackage{amsmath}
\usepackage{comment}

\def\qual{5-$\sigma$} % name for ``high-quality'' astrometry
\def\kms{km/s}
\def\dr2{\textit{Gaia}~DR2}
\def\gaia{\textit{Gaia}}
\def\gid#1{\textit{Gaia}~DR2~{#1}}
\def\gidinlist#1{{#1}}
\def\dsun{{d}_\text{h}}
\def\dsuni{{d}_\text{h,i}}
\def\vesc{{v}_\text{esc}}
\def\vgc{{v}_\text{g}}
\def\vgcerr{{\sigma}_{v,\text{g}}}
\def\vgcvec{\vec{v}_\text{g}}
\def\rgc{{r}_\text{g}}
\def\rgcerr{{\sigma}_{r,\text{g}}}
\def\rgcvec{\vec{r}_\text{g}}
\def\vgcunit{\hat{v}_\text{g}}
\def\rgcunit{\hat{r}_\text{g}}
\def\vsungcvec{\vec{v}_{\odot,\text{g}}}
\def\rheliounit{\hat{r}}
\def\vrrv{v_\text{r,g}}
\def\vtpm{v_\text{t,g}}
\def\vtpmvec{\vec{v}_\text{pm,g}}
\def\vzvperp{{v_\text{Z}}/v_\bot}
\def\pub{p_{\text{unb}}}
\def\pbo{p_{\textit{b,out}}}
\def\pbx{p_{\textit{b}}}
\def\Nsamp{N_{\textit{s}}}
\def\Nbx{N_{\textit{b}}}

\def\pax{\varpi}
\def\rv{v_\text{r}}
\def\LZ{L_\text{Z}}

\def\pmvec{\vec{\mu}}
\def\paxerr{\sigma_{\pax}}
\def\pmra{\mu_{\alpha}}
\def\pmdec{\mu_{\delta}}
\def\bmag{G_\text{BP}} % math mode

\def\brcolor{\text{BP-RP}} 
\def\fourvecel{{{\xi}}}
\def\fourvec{{\vec{\fourvecel}}}

\def\Cov{{\cal C}}
\def\Pprior{P_{\text{prior}}}
\def\Ppost{P_{\text{post}}}
\def\Pcond{P_{\text{cond}}}

\def\dupsrc{{\texttt{duplicated\_source}}}

\begin{document}

%\turnoffedit

\title{Nearby high-speed stars in Gaia DR2}

\author{Benjamin C. Bromley}
\affil{Department of Physics \& Astronomy, University of Utah, 
Salt Lake City, UT 84112}
\author{Scott J. Kenyon}
\affil{Smithsonian Astrophysical Observatory,
Cambridge, MA 02138}
\author{Warren R. Brown}
\affil{Smithsonian Astrophysical Observatory,
 Cambridge, MA 02138}
\author{Margaret J. Geller}
\affil{Smithsonian Astrophysical Observatory,
 Cambridge, MA 02138}

\begin{abstract}

We investigate the nature of nearby (10--15~kpc) high-speed stars in
the \dr2\ archive identified on the basis of parallax, proper motion
and radial velocity. Together with a consideration of their kinematic,
orbital, and photometric properties, we develop a novel strategy for
evaluating whether high speed stars are statistical outliers of the
bound population or unbound stars capable of escaping the Galaxy.  Out
of roughly 1.5 million stars with radial velocities, proper motions,
and \qual\ parallaxes, we identify just over 100 high-speed stars.  Of
these, only two have a nearly 100\% chance of being unbound, with
indication that they are not just bound outliers; both are likely
hyper-runaway stars.  The rest of the high speed stars are likely
statistical outliers.  We use the sample of high-speed stars to
demonstrate that radial velocity alone provides a poor discriminant of
nearby, unbound stars.  However, these stars are efficiently
identified from the tangential velocity, using just parallax and
proper motion. Within the full \dr2\ archive of stars with
\qual\ parallax and proper motion but no radial velocity, we identify
a sample of 19 with speeds significantly larger than the local escape
speed of the Milky Way based on tangential motion alone.

\end{abstract}


\section{INTRODUCTION}

%request to cite: https://arxiv.org/abs/1509.03633
%http://adsabs.harvard.edu/abs/2015A\%26A...584A..11L

The recent release of the \dr2\ catalog \citep{gaia2018a} has renewed
interest in the highest velocity stars in the Galaxy. In addition to
contributing new samples of candidates
\citep[e.g.,][]{marchetti2018b,shen2018,hattori2018b}, \gaia\ proper
motions and radial velocities allow more robust assessments of the
distances and space velocities of previously identified high velocity
stars \citep[e.g.,][]{boubert2018,brown2018,raddi2018}. As a result of
these analyses, some stars are clearly unbound. Others are just barely
bound to the Milky Way.

The 6-D positions and velocities available for over 7 million sources
in the \dr2\ archive also enable new tests of theoretical models for
the highest velocity stars. In the current paradigm, hyper-runaway
stars (HRSs) result from dynamical interactions among groups of
massive stars \citep[e.g.,][]{pov1967,leon1991,perets2012} or the
explosion of a massive star in a close binary
\citep[e.g.,][]{blaauw1961,dedonder1997, port2000,perets2012}.  The
supermassive black hole in the Galactic Center may also disrupt a
close binary system, capture one component, and eject the other as a
hypervelocity star \citep[HVS;][]{hills1988,yu2003}. Other physical
mechanisms may also accelerate stars to high velocity
\citep[e.g.,][]{sesana2006,yu2007,sesana2009,abadi2009,piffl2014,
  capuzzo2015,fragione2016,subr2016,hamers2017}.  Comparisons between
observed and predicted space motions yield constraints on the Galactic
potential and the ejection mechanism
\citep[e.g.,][]{bromley2006,kenyon2008,bromley2009,
  kenyon2014,rossi2014,rossi2017,hattori2018a,marchetti2018a,kenyon2018}.

Here, we use the \dr2\ proper motion and radial velocity data for the
brightest stars to test several aspects of theoretical models for HRSs
and HVSs \citep[see also][]{marchetti2018a,marchetti2018b}. Based on
existing samples of HVSs with B-type spectra in the outer halo
\citep[e.g.,][]{brown2005, edel2005, hirsch2005, brown2006a,
  brown2006b, brown2007a, brown2007b, brown2009b, brown2012a, brown2013, 
  brown2014, brown2015a, brown2015b}, the probability of detecting more 
than one B-type HVS within 10~kpc of the Sun is small \citep{kenyon2008,
  kenyon2014}.  Finding nearby HRSs is much easier \citep{bromley2009,
  kenyon2014}.  For either HRSs or HVSs, surveys using the Galactic
rest-frame tangential velocity should return a higher proportion of
nearby high velocity stars than the radial velocity
\citep{kenyon2018}. Our goal is to test these predictions with
\gaia\ data.

Identifying true high-velocity outliers in the \gaia\ DR2 archive is
challenging.  Among the roughly 7 million stars with measured
parallax, proper motion, and radial velocity, no more than a few
hundred candidates emerge with Galactic rest-frame velocity close to
or exceeding the local escape velocity
\citep[e.g.,][]{marchetti2018b,hattori2018b}.  Based on the quoted
errors of the measured quantities, only a few outliers unambiguously
exceed the escape velocity. Our goal is to consider the wealth of
kinematic information available in the \gaia\ data to analyze the
distribution of dynamical parameters and their errors and to make
robust estimates of the number of true outliers in the velocity
distribution.

Aside from identifying potential HRSs or HVSs, our analysis is important
for measuring the local escape velocity and the mass of the Milky Way
\citep[e.g.,][]{patel2018, gaia2018mw, watkins2018, posti2018,
  monari2018}.  All of the \gaia\ stars with radial velocity data lie
within roughly 15~kpc of the Sun. Understanding which of these stars
have velocities smaller than the local escape velocity helps to
establish the mass of the Milky Way within 20~kpc of the Galactic
Center \citep[see also][]{monari2018}.  Our analysis provides a
strategy to isolate unbound outliers from those bound to the Milky
Way.

We begin with a discussion of the sample selection in
\S\ref{sec:sampsel} and the basic properties of candidate high
velocity stars in \S\ref{sec:highspeed}. In \S\ref{sec:getvel}, we
assess the effectiveness of using radial velocities and proper motion
separately to identify high-speed sources, and in \S\ref{sec:newvtpm}
we introduce a new set of candidate stars selected on the basis of
proper motion only. Comparisons with theory follow in
\S\ref{sec:theory}. We conclude with a brief summary in
\S\ref{sec:conclusion}.

\section{SAMPLE SELECTION}\label{sec:sampsel}

We select stars with well-constrained position and velocity vectors,
seeking sources with measurements of radial velocity $\rv$, parallax 
$\pax$, and proper motion $\pmvec = (\pmra,\pmdec)$, where the
vector components are along right ascension $\alpha$ and declination
$\delta$, respectively. The \dr2\ archive \citep{gaia2016} lists
7,224,631 stars with estimates of all of these parameters.  This
section describes our astrometric selection of these ``6-D'' stars,
distance estimation and a novel numerical approach to it, and the
measurement of source parameters in the Galactic rest frame.

Nearly all stars with high-quality astrometry and radial
velocity data are bound to the Galaxy. Defining $\pub$ as the
probability a star is unbound, we seek stars with a 3-$\sigma$
confidence level $\pub \ge$ 0.997. To construct a reasonable set of
outliers, we identify groups of stars with $\pub \ge$ 0.2 and $\pub
\ge$ 0.5 \citep[see also][]{marchetti2018b,hattori2018b}. In
\S\ref{sec:getvel}, we develop a new approach to quantify this
probability based on a consideration of the error distribution for the
large population of bound stars.  This method enables us to determine
whether an unbound candidate has measured kinematical properties that
stand out from the bound outliers. A handful of the highest-speed
stars emerge as promising.

\subsection{Astrometric selection criteria}
%\subsection{Sample selection}

To select stars with high-quality astrometry, we follow the
recommendations in the basic source parameter descriptions
\citep{gaia2018} by requiring
\begin{align}
\nonumber
 &\texttt{astrometric\_gof\_al}<3,
&
&\texttt{astrometric\_excess\_noise} \leq 2, \\ 
\nonumber
 -0.23< &\texttt{mean\_varpi\_factor\_al} \leq 0.32,
& 
&\texttt{visibility\_periods\_used} > 9.
\end{align} 
We also make cuts based on photometry that impact the quality
of the astrometry:
\begin{eqnarray}
\nonumber
&
1+0.015 \times (\texttt{bp\_rp})^2 < \texttt{phot\_bp\_rp\_excess\_factor}
< 1.3+0.06 \times (\texttt{bp\_rp})^2
&
\\
\nonumber
& \chi_\nu^2 <
   1.2 \times \max\{1.0,\exp[-0.2*(\texttt{phot\_g\_mean\_mag}-19.5)]\}
&
\\ 
\nonumber
& \chi_\nu^2 \equiv 
\texttt{astrometric\_chi2\_al}/(\texttt{astrometric\_n\_good\_obs\_al}-5)
&
\end{eqnarray} 
\citep[see][Equations C.1 and C.2, therein]{lindegren2018}. 
In addition, we admit only those sources that have 
\begin{eqnarray}
\nonumber 
  &  \texttt{rv\_nb\_transits} > 5, & 
\end{eqnarray}
a condition that provides assurance of the quality of the reported
radial velocity, indicating that $\rv$ measurements were taking at a
minimum of six distinct epochs. This step helps to eliminate confusion
from binary stars. Applying all of these criteria yields 1,519,451 stars.

The basic source parameter, \dupsrc, indicates 
possibly compromised astrometry for stars in crowded fields 
\citep{gaia2018}. We accept stars even when this flag is raised
(about 13\% of the roughly $1.5\times 10^6$ objects) to enable
comparisons with previous results that do not use this selection 
criterion \citep[see][]{marchetti2018b,hattori2018b}. For the 
high-speed objects of primary interest, this flag is raised once 
in a set of 25 objects.  

Here, we include only stars with median parallax $\pax > 0$ and
relative error $\paxerr/\pax < 0.2$, corresponding to a 5-$\sigma$
detection. This threshold enables a straightforward estimate of
distance from the inverse of the parallax \citep[e.g.,][although see
  \citealt{luri2018} and our next discussion]{bailer-jones2015}. It is
also restrictive enough to provide meaningful estimates of whether
sources are unbound to the Galaxy, yet it eliminates fewer than
3\%\ of the sources that survive the \gaia-recommended quality cuts
listed above. The result is a ``\qual\ sample'' of $\Nsamp=$1,475,207
stars.

\subsection{Distance estimates from \gaia\ parallaxes}

A first step in assessing how sources in our sample are traveling with
respect to the Galaxy is the transformation of basic archive data into
physical distances and speeds.  A key element in the process is the
inference of heliocentric distance $\dsun$ from parallax. While the
significance of the parallax detections in our \qual\ sample is high,
parallax errors are not negligible. The parallax
error distribution of \dr2\ sources is well-approximated by a Gaussian
\citep{lindegren2018}, with a tail formally extending to
unphysical, negative values. When converting to distance using the
inversion formula $\dsun = 1/\pax$, even small values allowed by the
uncertainties may be unrealistic, based on the understanding that the
source is a star in the Milky Way.

A way to mitigate the problem is to select only sources that have
small relative parallax errors \citep[e.g.,][]{hattori2018b}.  When the
parallax error distribution is narrow compared to the measured
parallax, the extreme tails of the distribution are negligible. Then
the inverse of the median parallax gives a reliable distance estimate,
$\dsun = 1/\pax$. Inverted samples of the parallax error distribution 
also give a good representative set of possible distance measurements.

A more general approach, Bayesian inference, incorporates prior
information about source locations \citep[e.g.,][]{bailer-jones2015,
  astraatmadja2016, luri2018}. From Bayes' Theorem, the posterior
distribution of distances, $P(\dsun|\pax,\paxerr)$, given a measured
parallax $\pax$ with uncertainty $\paxerr$, is
\begin{equation}\label{eq:bayes}
  \Ppost(\dsun|\pax,\paxerr) \sim \Pcond(\pax|\dsun,\paxerr) \Pprior(\dsun),
\end{equation}
where $\Pcond(\pax|\dsun,\paxerr)$ is the distribution of parallax
values given a distance $\dsun$ (here, a Gaussian with mean
$1/\dsun$), and $\Pprior(\dsun)$ is the prior distribution that
contains assumptions about where a source is
located. \citet{bailer-jones2015} introduces
\begin{equation}\label{eq:prior}
\Pprior \sim \dsun^2 \exp(-\dsun/L),
\end{equation}
corresponding to an exponential fall-off in density, with a most
probable source location at $2L$.  \citet{marchetti2017} set $L =
2.6$~kpc to represent bulk of the \gaia\ stars, averaged over the
plane of the sky.  \citet{bailer-jones2018} adopt values around 1~kpc,
depending on the sky location relative to the Galaxy.
However, we hesitate to adopt priors that are tailored to the bulk
catalog in a search for rare, unbound stars on orbits that are not
known \textit{a priori}.

When the parallax uncertainty is large compared to the measured
parallax, Bayesian distance estimates are dominated by the choice of
prior. When the relative parallax errors are small, reasonable priors
have little influence on the inferred distance. Our threshold,
$\paxerr/\pax = 0.2$, is on the boundary between these two cases
\citep[e.g.,][]{bailer-jones2015, astraatmadja2016}.  For a 5-$\sigma$
detection, nearly all parallaxes consistent with measurement of a
source correspond to physical distances, enabling a blind search for
rare, unknown sources. Furthermore, the \qual\ sample is not strongly
affected by potential systematic errors in parallax
\citep{lindegren2018}.  Nonetheless, priors as in
equation~(\ref{eq:prior}) may modestly affect the noisier sources in the
our sample \citep[cf.][]{marchetti2018b}. Thus, we perform our
analysis with and without priors to understand their impact.  Our
analysis demonstrates that our most promising high-speed sources and
our overall conclusions are not affected by this choice.

\subsection{Galactocentric quantities}

Our analysis of the \qual\ sample requires estimates of a source's
position and velocity relative to the Galactic Center. To obtain these
quantities, we assume that the Sun has a position vector
$(X_\odot,Y_\odot,Z_\odot) = (-8,0,0)$~kpc and a velocity vector
$(U_\odot,V_\odot,W_\odot) = (11.1,247.24,7.25)$~\kms\ \citep[e.g.,]
[]{schonrich2010} in a right-handed coordinate system ($X,Y,Z,U,V,W$)
where the origin is fixed at the Galactic Center
\citep[e.g.,][]{johnson1987}.  The Sun's velocity includes a
contribution of $V = 235$~\kms\ from the rotation of the Galactic disk
\citep[e.g.][]{reid2014}. These definitions allow us to transform
\gaia\ measurements into the Galactic frame of reference.

To estimate Galactocentric position, velocity, and derived quantities
from the parallax, proper motion, and radial velocity, we account for
uncertainties and correlations between measurements as specified in
\dr2\ basic source parameter list. We group together the measured quantities
as a vector,
\begin{equation}
  \fourvec = \{\fourvecel_i\} \equiv (\pax,\pmra,\pmdec,\rv),
\end{equation}
and, with index $i$ running from one to four, write the elements of the
corresponding covariance matrix,
\begin{equation}\label{eq:covel}
\Cov_{ii} = \sigma_i^2 \text{\ \ and\ \ } 
\Cov_{ij} = \sigma_i \sigma_j \rho_{ij}\ \ (i\neq j),
\end{equation}
where $\sigma_i$ is the uncertainty in $\fourvecel_i$, and $\rho_{ij}$
is the correlation coefficient for measurements $\fourvecel_i$ and
$\fourvecel_j$.  The \dr2\ archive lists all of the uncertainties and
correlation coefficients, except that radial velocity is uncorrelated
with parallax and proper motion, so $\rho_{i<4,4} = 0$.

We take a Monte Carlo approach to error estimation, generating samples
from the joint distribution of parallax, proper motion and radial
velocity, as specified by the measured median values, uncertainties
and correlations. In our ``Quasi-Monte Carlo'' implementation,
low-discrepancy sequences form realizations of quasi-random
4-vectors, uniformly distributed in the unit hypercube
\citep{sobol1976, press1992}. A pair of 2-D Box-Muller transformations
converts each quasi-random vector into a new 4-vector $\vec{n}$, whose
components are (quasi-)independent and normally distributed. A
Cholesky factorization of the covariance matrix $\Cov$ in
equation~(\ref{eq:covel}) transforms this vector into the
% @@ bcb fixed C^{-1/2}, should be C^{1/2}
\gaia\ observables, $\fourvec = \Cov^{1/2}\vec{n}$. In this way, we
generate $N=4,096$ realizations of $(\pax,\pmra,\pmdec,\rv)$ per
star. If heliocentric distance estimates come from $1/\pax$, then
samples are converted geometrically into desired quantities like
Galactocentric distance ($\rgc$) and speed ($\vgc$).

When heliocentric distance $\dsun$ comes from Bayesian estimation, the
Quasi-Monte Carlo procedure just described is modified to sample a 4-D
distribution with $\dsun$ replacing parallax as a variate, distributed
according to the posterior, $\Ppost$. To maintain correlations with
proper motion samples, practitioners recommend Markov Chain Monte
Carlo (MCMC) methods \citep[e.g.,][]{marchetti2017, luri2018}.
Because the distance priors considered here are 
% @@bcb
proper and 
functions of $\dsun$
alone, we instead calculate the cumulative posterior distribution,
$\Ppost(<\dsun)$, and numerically invert it. The $i$-th sample of a 
uniform variate $u$ then yields distance $\dsuni = \Ppost^{-1}(u_i)$. To
preserve correlations between $\dsuni$ and proper motion, we generate
Quasi-Monte Carlo trials as before
% @@bcb. old:
%but adjust the Gaussian distributed
%parallax samples to become appropriately ordered samples of $u$:
%\begin{equation}
%u_i \equiv 
%\frac{1-\text{erf}[(\pax_i-\pax)/\sqrt{2\paxerr^2}]}%
%{1+\text{erf}(\pax/\sqrt{2\paxerr^2})}%
%\ \ \ \ (\pax_i > 0).
%\end{equation}
% @@bcb. new:
but in the transformation of the $i$-th set of independent normal
quasi-random variates to observables, $\fourvec_{i} = \Cov^{1/2}\vec{n}_i$,
we make the substitution
\begin{equation} 
n_{1,i} \rightarrow (1/\dsuni - \pax)/\paxerr
\ \ \ \ (\dsuni > 0).
\end{equation}
This step takes advantage of the lower triangular form of $\Cov^{1/2}$
from the Cholesky factorization.
% @@bcb. done.
% @@bcb. also don't need:
%We perform this operation within the Quasi-Monte Carlo algorithm. 
As a check, we reproduce results obtained with a full 4-D MCMC solver from
the \texttt{emcee} package \citep{foreman-mackey2013}.

Compared to MCMC, and to pseudorandom methods generally, the
Quasi-Monte Carlo approach converges quickly, like a grid-based
integrator \citep[e.g.,][]{press1992}.  For example, with $N=4,096$,
estimates of $\vgcerr$, the 1-$\sigma$ error in Galactocentric speed,
are typically accurate to within a few tenths of a
percent. Pseudorandom Monte Carlo trials with the same $N$ yield
estimates with uncertainties of a few percent. The MCMC approach also
gives errors of several percent with that same number of trials, not
including the ``burn-in'' steps \citep{foreman-mackey2013}.

\subsection{Orbit selection: bound versus unbound}
\label{subsec:vesc}

Next we identify high-speed stars that are potentially on unbound
orbits relative to the Galaxy. We use the Galaxy model composed of a
compact bulge, a rotating disk, and an extended, dark matter halo
described in \citet{kenyon2014, kenyon2018}.  The Galactic potential
$\Phi$ is the sum of the potential of each component; the model
parameters \citep[][\S 3.1 therein]{kenyon2018} are consistent with
recent \gaia\ observations of the mass of the Milky Way
\citep[cf.][]{callingham2018, fritz2018, watkins2018}.

From the gravitational potential, we obtain a local escape speed,
$\vesc(\rgcvec)$, defined as the minimum speed of a star that reaches
a distance of 250~kpc from the Galactic Center:
\begin{equation}
\vesc = \sqrt{2[\Phi(\vec{r}_{250}) - \Phi(\rgcvec)]}.
\end{equation}
For simplicity, we take $\vec{r}_{250}$ as a point at that
distance on the symmetry axis of the Galactic disk.  An unbound orbit
has $\vgc \ge \vesc$. Quasi-Monte Carlo draws of $\rgc$ and $\vgc$ from
the error distributions of each source give the fraction of trials
that yield unbound orbits; this fraction is our estimator for $\pub$,
the probability that a star is unbound.

There are 25 candidate high velocity stars with $\pub$ exceeding 50\%.
An additional 101 candidates have a 20\% or better likelihood of being
on an unbound orbit ($\pub \ge 0.2$).  Table~\ref{table:hvseq}
reproduces the \dr2\ basic source parameters for the 25 high-speed
outliers.

% \begin{rotatetable}

\begin{deluxetable}{lcrrrrccl}
%\tabletypesize{\scriptsize}
\tabletypesize{\tiny}
\tablewidth{6.75in}
\tablecaption{High-speed stars: \dr2\ basic source parameters
\label{table:hvseq}}
\tablehead{%
\\
\colhead{\gaia\ DR2} & \colhead{($\alpha$, $\delta$)} &
\colhead{$\pax$} & \colhead{$\pmra$} & \colhead{$\pmdec$} &
\colhead{rv} & \colhead{G} &
 \colhead{$\bmag$} & 
\colhead{$\brcolor$}  
\\
\colhead{designation} & \colhead{(deg)} &
\colhead{(mas)} & \colhead{(mas/yr)} & \colhead{(mas/yr)} &
\colhead{(\kms)} & \colhead{(mag)} &
 \colhead{(mag)} & 
\colhead{(mag)}  
}
\startdata
5932173855446728064$^{M,d}$ & (244.118100,-54.440452) & 0.454$\pm$0.029 &  -2.68$\pm$0.04   &  -4.99$\pm$0.03   &  -614.29$\pm$2.49 &  13.81 &  14.21 & 0.99\\
1383279090527227264$^{M}$ & (240.337348, 41.166774) & 0.118$\pm$0.016 & -25.76$\pm$0.03   &  -9.75$\pm$0.04   &  -180.90$\pm$2.42 &  13.01 &  13.51 & 1.16\\
1478837543019912064 & (212.477709, 33.712932) & 0.105$\pm$0.019 & -17.61$\pm$0.02   & -16.57$\pm$0.03   &  -245.88$\pm$1.48 &  13.09 &  13.51 & 1.00\\
6456587609813249536$^{M}$ & (317.360892,-57.912400) & 0.099$\pm$0.019 &  13.00$\pm$0.03   & -18.26$\pm$0.03   &   -15.85$\pm$2.83 &  13.01 &  13.47 & 1.08\\
6492391900301222656$^{M}$ & (348.646647,-58.429575) & 0.095$\pm$0.018 &   7.50$\pm$0.03   & -15.82$\pm$0.03   &  -149.86$\pm$1.16 &  13.36 &  13.94 & 1.29\\
4326973843264734208$^{M}$ & (248.892295,-14.518435) & 0.199$\pm$0.031 & -20.55$\pm$0.05   & -33.97$\pm$0.03   &  -220.39$\pm$2.05 &  13.50 &  14.43 & 1.87\\
5846998984508676352 & (211.089783,-69.345114) & 0.095$\pm$0.019 & -16.20$\pm$0.02   &  -2.69$\pm$0.03   &    31.43$\pm$4.38 &  14.14 &  14.92 & 1.63\\
2089995308886282880$^{M}$ & (280.928177, 31.345968) & 0.071$\pm$0.013 &  -3.17$\pm$0.02   &  -9.50$\pm$0.02   &    -9.90$\pm$0.52 &  13.18 &  13.89 & 1.51\\
5802638672467252736$^{M}$ & (255.717030,-74.057467) & 0.101$\pm$0.015 &  -8.72$\pm$0.02   & -15.29$\pm$0.02   &    70.53$\pm$1.46 &  13.06 &  13.70 & 1.39\\
2095397827987170816$^{M}$ & (276.654116, 35.056068) & 0.066$\pm$0.012 &  -3.13$\pm$0.02   &  -8.86$\pm$0.02   &   -96.95$\pm$1.11 &  13.45 &  14.15 & 1.51\\
6431596947468407552$^{M}$ & (274.687922,-70.249323) & 0.084$\pm$0.016 &   4.55$\pm$0.02   &   4.97$\pm$0.02   &   259.08$\pm$1.65 &  13.09 &  13.66 & 1.28\\
2159020415489897088$^{M}$ & (273.321443, 61.318680) & 0.134$\pm$0.026 &   3.99$\pm$0.05   &  15.67$\pm$0.05   &  -162.28$\pm$0.99 &  12.51 &  13.12 & 1.33\\
5919596571872806272 & (265.896592,-56.104065) & 0.120$\pm$0.022 & -11.03$\pm$0.04   & -20.09$\pm$0.04   &   212.10$\pm$2.21 &  13.01 &  13.57 & 1.28\\
2121857472227927168$^{M}$ & (275.124461, 47.497863) & 0.072$\pm$0.013 &  -5.36$\pm$0.02   &  -7.06$\pm$0.02   &  -434.70$\pm$0.71 &  13.27 &  13.96 & 1.49\\
5839686407534279808$^{M}$ & (209.437107,-72.149655) & 0.138$\pm$0.020 & -22.74$\pm$0.03   &  -3.18$\pm$0.03   &   175.38$\pm$4.87 &  13.91 &  14.84 & 1.87\\
2112308930997657728$^{M}$ & (272.894471, 39.889802) & 0.167$\pm$0.022 & -21.92$\pm$0.04   & -12.60$\pm$0.04   &  -107.02$\pm$1.40 &  12.58 &  13.02 & 1.04\\
6656557095228727936 & (286.480891,-52.679280) & 0.105$\pm$0.020 &   2.31$\pm$0.03   & -18.83$\pm$0.02   &   -85.25$\pm$2.60 &  13.43 &  13.94 & 1.17\\
5399966178291369728$^{M}$ & (166.880803,-37.647268) & 0.100$\pm$0.017 & -12.87$\pm$0.02   &   0.51$\pm$0.02   &   420.38$\pm$1.84 &  13.08 &  13.69 & 1.36\\
4366218814874247424$^{M}$ & (256.061406, -2.675249) & 0.139$\pm$0.021 & -19.29$\pm$0.04   &   6.03$\pm$0.03   &  -132.81$\pm$1.22 &  13.17 &  13.89 & 1.54\\
5217818333256869376$^{M}$ & (141.829543,-73.543300) & 0.118$\pm$0.018 & -13.27$\pm$0.04   &   8.27$\pm$0.04   &   375.09$\pm$1.26 &  12.48 &  13.10 & 1.37\\
6124121132097402368 & (212.016640,-32.476408) & 0.120$\pm$0.024 & -19.81$\pm$0.04   &  -2.60$\pm$0.04   &    85.24$\pm$1.21 &  13.05 &  13.61 & 1.25\\
2106519830479009920$^{M}$ & (285.484415, 45.971657) & 0.123$\pm$0.018 &   3.30$\pm$0.04   &  13.17$\pm$0.04   &  -212.12$\pm$0.98 &  12.42 &  13.04 & 1.35\\
5835015235520194944 & (244.519790,-58.328708) & 0.118$\pm$0.020 & -16.31$\pm$0.03   & -13.74$\pm$0.03   &   106.78$\pm$1.49 &  13.17 &  13.99 & 1.70\\
1989862986804105344$^{M}$ & (340.509702, 51.611096) & 0.095$\pm$0.016 &  10.34$\pm$0.02   &   5.09$\pm$0.02   &   -75.70$\pm$1.62 &  13.08 &  13.72 & 1.39\\
5779919841659989120$^{M}$ & (235.357736,-77.283183) & 0.094$\pm$0.016 & -11.43$\pm$0.03   &  -8.40$\pm$0.03   &   -13.31$\pm$0.83 &  13.55 &  14.39 & 1.71
\enddata
%\tablecomments{The first object has the highest radial velocity, 
\tablecomments{\parbox[t]{6in}{This list is in order of decreasing
probability of being unbound (see Table~\ref{table:hvsgc}).
    In the first column, the superscript ``d'' indicates
    that the \dupsrc\         
    flag is raised, while the ``M'' indicates
    that a source is also listed in \citet{marchetti2018b}.}}
%corrected for Solar motion. The remaining }
\end{deluxetable}
%\end{rotatetable}

Table \ref{table:hvsgc} lists Galactocentric data for the 25 high-speed 
outliers, sorted by decreasing probability of being unbound. 
It includes heliocentric distance ($\dsun$), Galactocentric distance 
($\rgc$) and speed ($\vgc$), along with Galactocentric speeds inferred 
from either the heliocentric radial velocity or proper motion:
\begin{eqnarray}
\label{eq:vrrv}
\vrrv & = & \rv + \vsungcvec \cdot \rheliounit,
\\ 
\label{eq:vtpm}
\vtpmvec & = & k\frac{\pmvec}{\pax} 
+ \vsungcvec-(\vsungcvec\cdot\rheliounit) \rheliounit,
\end{eqnarray}
where $\vsungcvec$ is the Sun's velocity in the Galaxy's rest frame,
and $\rheliounit$ is the unit vector in the direction of the Sun in
that frame.  Equation~(\ref{eq:vrrv}) gives the observer-frame
line-of-sight component of the Galactic velocity, corrected for Solar
motion; equation (\ref{eq:vtpm}) gives the Galactocentric velocity
vector in the observer's sky plane, corrected for Solar motion, as
estimated from the parallax and the proper motion vector $\pmvec =
(\pmra,\pmdec)$ in equatorial coordinates.  The constant $k$ in
equation (\ref{eq:vtpm}) has a value of 4.7047 when speed, parallax,
and proper motion are in units of \kms\, mas, and mas/yr,
respectively.

Table~\ref{table:hvsgc} also includes $\gamma$, the angle between each 
star's radial position and velocity:
\begin{equation}
\cos \gamma = \vgcunit\cdot\rgcunit =
 \frac{\rgcvec \cdot \vgcvec}{|\rgcvec||\vgcvec|}.
\end{equation}
Stars on purely radial, outbound orbits have $\gamma = 0^\circ$; stars
orbiting the Galactic Center have $\gamma = 90^\circ$.  The $\LZ/L_{200}$
parameter, the $Z$-component of the stars' angular momentum relative 
to a disk-like orbit with $|\vec{L}_{200}| = \rgc \times 200$~\kms,
indicates how stars move relative to the rotation of stars in the 
Milky Way's disk. Similarly, the quantity
\begin{equation}
  \vzvperp \equiv\ \frac{W}{\left(U^2+V^2\right)^{1/2}}
\end{equation}
establishes out-of-plane motion as compared with the speed in a 
plane parallel to the disk. This measure also serves to distinguish 
disk stars from the halo population.

%\begin{rotatetable}
\begin{deluxetable}{lcrrrrrrrrc}%
%\tabletypesize{\scriptsize}
\tabletypesize{\tiny}
\tablecaption{High-stars stars: Galactocentric kinematical parameters.
\label{table:hvsgc}}
\tablewidth{7.0in}
\tablehead{%
\\
\colhead{\gaia\ DR2} & \colhead{($\ell$,$b$)} &
\colhead{$\dsun$ } & 
\colhead{$\rgc$ } & \colhead{$\vgc$} &
\colhead{$\vrrv$} & \colhead{$\vtpm$ } &  \colhead{$\gamma$} & 
 \colhead{$\LZ/L_{200}$} & 
\colhead{$\vzvperp$} &
\colhead{$\pub$} 
\\ 
\colhead{designation} & \colhead{(deg)} &
\colhead{(kpc)} & \colhead{(kpc)} & \colhead{(\kms)} &
\colhead{\kms} & \colhead{\kms} &  \colhead{(deg)} & 
 \colhead{\ } & 
\colhead{\ } & 
\colhead{\ } 
}
\startdata
5932173855446728064 & (329.9, -2.7) &  2.2$\pm$0.1  &  6.2$\pm$0.1  &  747$\pm$3   & -728.7$\pm$2.5   &  164$\pm$4   &  53.0$\pm$0.6  &  3.0$\pm$0.0 &  0.02$\pm$0.00 & 1.00 \\
1383279090527227264  & ( 65.5, 48.8) &  8.5$\pm$1.3  & 10.0$\pm$0.9  &  924$\pm$168 &  -24.4$\pm$2.4   &  924$\pm$168 &  86.7$\pm$0.6  & -2.3$\pm$0.3 &  0.79$\pm$0.01 & 1.00 \\
1478837543019912064  & ( 59.0, 71.9) &  9.6$\pm$2.1  & 11.4$\pm$1.6  &  876$\pm$230 & -171.3$\pm$1.5   &  859$\pm$234 &  95.9$\pm$0.7  & -2.5$\pm$0.5 &  0.09$\pm$0.04 & 0.99 \\
6456587609813249536  & (338.3,-40.9) & 10.1$\pm$2.4  &  7.3$\pm$1.7  &  889$\pm$250 &  -82.0$\pm$2.8   &  885$\pm$251 &  51.3$\pm$9.0  &  0.8$\pm$2.5 & -0.33$\pm$0.01 & 0.98 \\
6492391900301222656  & (324.6,-54.4) & 10.5$\pm$2.4  &  9.7$\pm$1.8  &  678$\pm$189 & -233.8$\pm$1.2   &  637$\pm$198 &  93.0$\pm$0.4  & -0.3$\pm$0.7 &  0.60$\pm$0.09 & 0.81 \\
4326973843264734208  & (  2.6, 21.5) &  5.0$\pm$0.9  &  3.8$\pm$0.4  &  730$\pm$159 & -197.0$\pm$2.1   &  703$\pm$164 &  91.8$\pm$5.3  & -3.1$\pm$0.4 & -0.24$\pm$0.03 & 0.79 \\
5846998984508676352  & (309.3, -7.4) & 10.5$\pm$2.5  &  8.3$\pm$1.9  &  671$\pm$189 & -152.2$\pm$4.4   &  654$\pm$193 &  55.9$\pm$6.9  &  2.6$\pm$3.9 &  0.15$\pm$0.01 & 0.74 \\
2089995308886282880  & ( 60.7, 15.2) & 14.2$\pm$3.2  & 12.6$\pm$2.8  &  605$\pm$144 &  205.3$\pm$0.5   &  569$\pm$151 & 104.3$\pm$2.7  &  2.8$\pm$1.9 & -0.07$\pm$0.00 & 0.73 \\
5802638672467252736  & (317.9,-19.1) &  9.9$\pm$1.6  &  7.2$\pm$1.1  &  648$\pm$135 &  -80.6$\pm$1.5   &  643$\pm$136 &  45.6$\pm$7.8  &  1.9$\pm$1.2 & -0.10$\pm$0.01 & 0.70 \\
2095397827987170816  & ( 63.0, 19.9) & 15.1$\pm$3.3  & 13.7$\pm$2.9  &  585$\pm$142 &  117.4$\pm$1.1   &  573$\pm$144 & 110.3$\pm$3.7  &  2.5$\pm$1.7 & -0.05$\pm$0.01 & 0.69 \\
6431596947468407552 & (324.2,-22.7) & 12.0$\pm$2.7  &  8.0$\pm$2.2  &  605$\pm$84  &  131.2$\pm$1.7   &  591$\pm$85  & 105.4$\pm$5.7  & -1.7$\pm$1.5 & -0.44$\pm$0.00 & 0.66 \\
2159020415489897088  & ( 90.5, 28.1) &  7.5$\pm$1.8  & 11.0$\pm$1.3  &  588$\pm$134 &   59.2$\pm$1.0   &  585$\pm$135 &  41.1$\pm$5.4  & -1.1$\pm$1.1 & -0.25$\pm$0.02 & 0.65 \\
5919596571872806272  & (336.2,-13.5) &  8.3$\pm$1.8  &  3.8$\pm$0.9  &  684$\pm$188 &  123.3$\pm$2.2   &  672$\pm$190 &  26.8$\pm$8.3  &  0.2$\pm$2.7 & -0.06$\pm$0.01 & 0.63 \\
2121857472227927168  & ( 75.5, 24.7) & 14.0$\pm$3.0  & 14.4$\pm$2.6  &  550$\pm$114 & -211.6$\pm$0.7   &  507$\pm$122 & 141.8$\pm$8.8  &  1.1$\pm$1.0 &  0.07$\pm$0.06 & 0.62 \\
5839686407534279808  & (308.0, -9.9) &  7.2$\pm$1.1  &  6.8$\pm$0.5  &  626$\pm$123 &  -11.0$\pm$4.9   &  625$\pm$123 &  25.7$\pm$6.7  &  1.0$\pm$0.8 &  0.12$\pm$0.00 & 0.62 \\
2112308930997657728  & ( 67.0, 24.2) &  6.0$\pm$0.8  &  8.1$\pm$0.4  &  601$\pm$100 &  107.4$\pm$1.4   &  592$\pm$101 & 118.0$\pm$1.5  &  0.3$\pm$0.2 &  0.96$\pm$0.02 & 0.62 \\
6656557095228727936  & (344.2,-23.4) &  9.5$\pm$2.1  &  4.5$\pm$1.5  &  661$\pm$187 & -140.2$\pm$2.6   &  646$\pm$190 &  50.7$\pm$13.3 &  1.3$\pm$2.9 & -0.41$\pm$0.01 & 0.62 \\
5399966178291369728  & (281.2, 20.8) & 10.0$\pm$2.0  & 11.7$\pm$1.6  &  564$\pm$117 &  198.2$\pm$1.8   &  528$\pm$124 &  31.2$\pm$8.5  &  0.9$\pm$1.0 & -0.11$\pm$0.05 & 0.59 \\
4366218814874247424  & ( 17.5, 22.3) &  7.2$\pm$1.2  &  3.8$\pm$0.3  &  651$\pm$108 &  -51.7$\pm$1.2   &  649$\pm$108 &  33.9$\pm$7.1  & -0.8$\pm$0.2 &  1.92$\pm$0.11 & 0.59 \\
5217818333256869376  & (289.7,-16.2) &  8.5$\pm$1.5  &  9.6$\pm$1.0  &  578$\pm$105 &  153.1$\pm$1.3   &  558$\pm$109 &  26.9$\pm$6.4  &  1.1$\pm$0.8 & -0.47$\pm$0.04 & 0.58 \\
6124121132097402368  & (321.1, 27.7) &  8.3$\pm$2.0  &  6.4$\pm$1.2  &  619$\pm$188 &  -41.1$\pm$1.2   &  618$\pm$189 &  30.5$\pm$12.2 &  1.1$\pm$2.1 &  0.32$\pm$0.04 & 0.58 \\
2106519830479009920  & ( 76.2, 17.4) &  8.1$\pm$1.3  & 10.0$\pm$0.9  &  567$\pm$82  &   21.7$\pm$1.0   &  566$\pm$82  &  37.0$\pm$5.4  & -1.6$\pm$0.7 &  0.01$\pm$0.01 & 0.56 \\
5835015235520194944  & (327.4, -5.6) &  8.5$\pm$1.6  &  4.7$\pm$0.9  &  647$\pm$166 &  -17.3$\pm$1.5   &  646$\pm$166 &  27.7$\pm$11.7 &  1.2$\pm$1.9 &  0.10$\pm$0.01 & 0.56 \\
1989862986804105344  & (103.4, -6.3) & 10.5$\pm$2.1  & 14.6$\pm$1.8  &  531$\pm$108 &  160.0$\pm$1.6   &  506$\pm$112 &  40.3$\pm$7.0  & -1.7$\pm$1.0 & -0.02$\pm$0.01 & 0.55 \\
5779919841659989120  & (311.9,-17.5) & 10.6$\pm$2.0  &  8.3$\pm$1.5  &  571$\pm$128 & -183.9$\pm$0.8   &  541$\pm$134 &  62.0$\pm$5.0  &  2.2$\pm$1.3 &  0.08$\pm$0.01 & 0.50 
\enddata
\tablecomments{\parbox[t]{6in}{The variables are described in the
    text, and illustrated in Figure~\ref{fig:gcvelgc}.}}
\end{deluxetable}

\subsection{Comparison with other work}

Our sample selection criteria, based on the quality of astrometry and
the probability of being unbound to the Galaxy, generates stars that
overlap those previously identified in detailed analyses of
\gaia\ data.  \citet{hattori2018b} identify 30 stars with complete
velocity measurements on the basis of high tangential speed relative
to the Galactic Center, and use low parallax uncertainty,
$\paxerr/\pax < 0.1$, as a surrogate for astrometric quality
cuts. Four of these stars are in our group of 101 marginally bound
stars; none of our other candidates satisfy the condition of low
parallax uncertainty adopted by \citet{hattori2018b}.  Of the
remaining 26 stars in their sample, we exclude 16 because
\gaia\ astrometric parameters (e.g., \texttt{astrometric\_gof\_al})
indicate possible problems with the parallax measurements. The other
10 stars are excluded in our sample on the basis of low unbound
probability, resulting from our choice of a more massive Milky Way
(MW) model with higher escape speeds.  \citet{hattori2018b} adopt the
\texttt{MWPotential2014} model of \citet{bovy2015}, which has a mass
that is 80\%\ of the \citet{kenyon2018} model.  Compared to a value of
578~\kms\ for the more massive MW, the less massive MW has a much
lower escape speed of $\vesc = 513$~\kms\ at $\rgc = 8$~kpc.  Aside
from yielding a more stringent assessment of the unbound probability,
the more massive MW model is favored by recent observations
\citep[e.g.,][]{watkins2018}.

\citet{marchetti2018b} also mine the \dr2\ stars with 3-D velocity
measurements, using nearly identical criteria as we do here
by following recommendations in \citet{gaia2018}.
%\footnote{The only
%  difference in the astrometric selection between
%  \citet{marchetti2018b} and this work is that we follow the
%  recommendation in \citet{gaia2018} to use
%  $\texttt{visibility\_periods\_used} > 9$ instead of 8.}  
However, they do not impose a cut based on parallax uncertainty, and
instead select sources with $\vgcerr/\vgc < 0.3$.  Their catalog of
125 unbound stars ($\pub \ge 0.5$) has 19 stars in common with our
list.  The remaining stars are excluded here because the parallaxes
(including $\pax<0$) do not meet our astrometric cuts for the
\qual\ sample (90 sources), or the $\pub$ estimates are higher than
ours because of the adopted Milky Way potential (16 sources). As in
\citet{hattori2018b}, \citet{marchetti2018b} use a model of the Milky
Way that is less massive than the one chosen here.  Our sample includes six
sources that do not make the \citet{marchetti2018b} criteria for
Galactocentric velocity uncertainty (see Table~1).

\citet{du2018} analyze 16 stars from \gaia\ DR 1 with spectroscopic 
data from LAMOST.  None of these objects are in our sample of unbound 
candidates; however, a single star,
\gid{3266449244243890176}, appears in our group of 101 marginally
bound candidates. Of the 16 stars in \citet{du2018}, 11 do not have
radial velocity data in the \dr2\ archive; four others with measured
$\rv$ have goodness-of-fit parameters \texttt{astrometric\_gof\_al}
outside of the recommended range that indicates reliable astrometry in
\gaia\ DR 2.

\section{THE HIGH-SPEED OUTLIERS}\label{sec:highspeed}

Here we examine characteristics of the 25 stars in our set of
high-speed outliers to address their evolutionary state and likely
origin.

\subsection{Photometric properties}

\begin{figure}
\centerline{\includegraphics[width=5in]{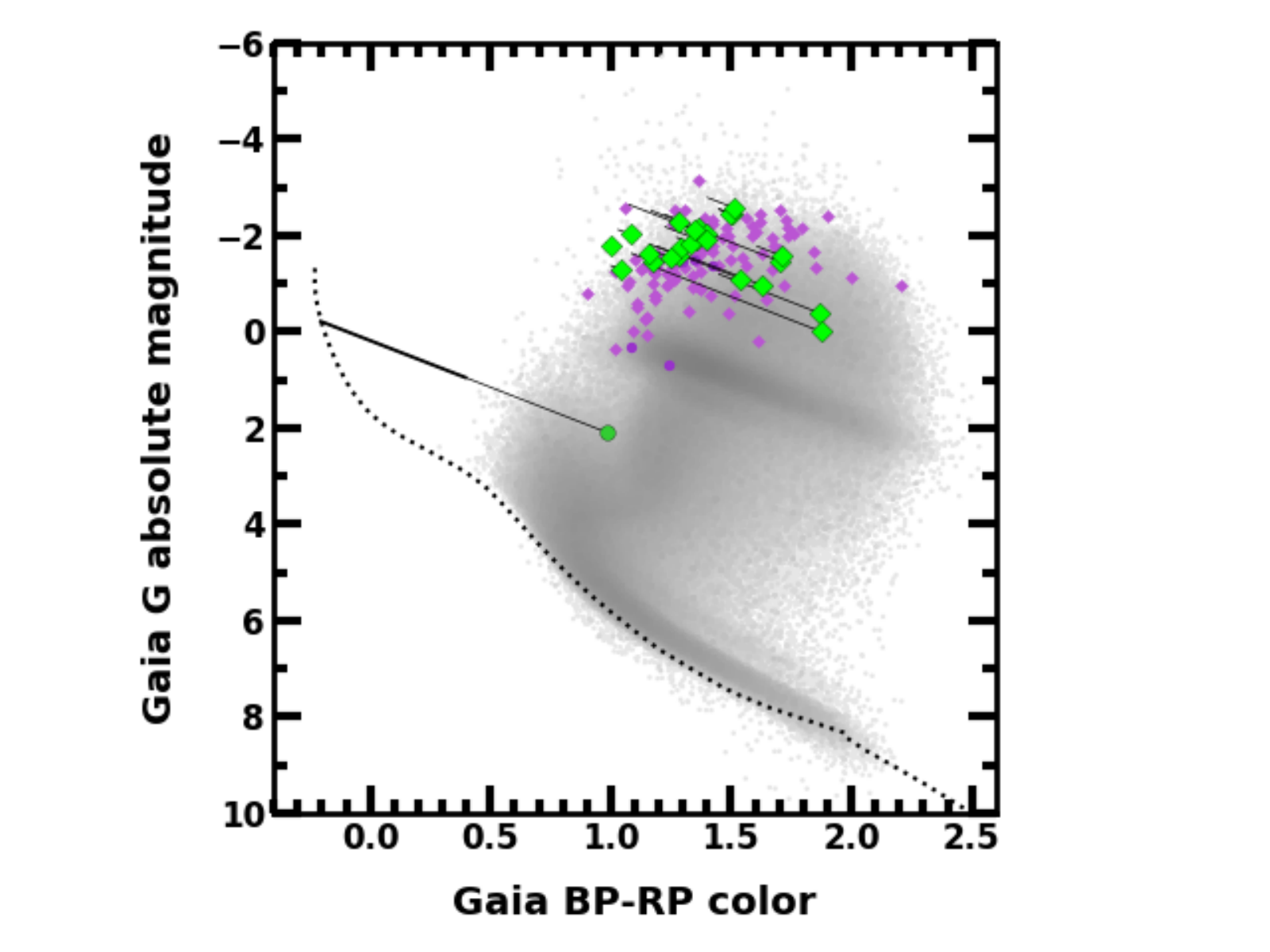}}
\caption{\label{fig:gcvelhr} Color-magnitude diagram for stars in the
  \dr2\ archive with \qual\ parallaxes and complete velocity
  measurements. The light gray points and shaded region show the
  distribution of the $\Nsamp \approx 1.5$-million sources with these
  properties (the shading is proportional to the logarithm of the
  density of stars in the plot). Main sequence stars are well defined
  on the lower left edge; giants lie above them.  The magenta diamonds
  indicate stars that are marginally bound ($0.2\leq \pub<0.5$); the
  lime-green diamonds are the unbound candidates ($\pub\geq 0.5$).
  The darker shade of green or magenta circles indicate sources that
  have the \dupsrc\ flag raised. For the unbound candidates, we also
  correct for reddening as described in text; the tips of the thin
  black lines indicate the estimated intrinsic color and brightness.
  The faintest and bluest unbound star, located toward the Galactic
  Center, has intrinsic color and magnitude that suggests it is on the
  main sequence (indicated by the dotted curve, based on
  \citealt{marigo2017} for a 100-Myr isochrone with a metallicity of
  0.01520). The thicker black line indicates a range of reddening and
  extinction corrections, with the \citet{marshall2006} dust map
  giving the smaller correction and the \citet{schlafly2011} estimate
  giving the upper limit.}
\end{figure}

Although we do not have optical spectroscopy for these stars, the
color-magnitude diagram in Figure~\ref{fig:gcvelhr} demonstrates that
most stars are late-type giants.  The plot shows \gaia\ G band
absolute magnitude, $M_G$, versus \gaia\ Blue-Red Photometer color,
$\brcolor$, without correction for Galactic reddening, for the full
sample with \qual\ parallaxes and radial velocity data.  The population
shows F-type and later stars on the main sequence and a set of late-type
giants.  Most of the fast moving stars are the brightest among the 
giants, well above the red clump of helium-burning, solar-type stars 
\citep{gaiahrd2018}.  Possibly, they may be younger (400~Myr--1~Gyr) 
and more massive stars in a similar evolutionary stage, members of 
a vertical red clump \citep{caputo1995, zaritsky1997, ibata1998}.  
The analysis of \citet[\S3.1 therein]{hattori2018b} suggests that 
these fastest stars are less massive, older ($\gtrsim$1~Gyr) 
metal-poor stars on the asymptotic giant branch.

To estimate reddening corrections for the 25 unbound candidates, we
use the IPAC/IRSA DUST
service\footnote{https://irsa.ipac.caltech.edu/applications/DUST},
based on the work of \citet{schlegel1998} and
\citet{schlafly2011}. With a default \texttt{Image Size} of 5~degrees,
we obtain the color excess, $E(B-V)$, and optical extinction, $A_V
= 3.1 E(B-V)$, for each star.  To convert to \gaia\ passbands, we
follow the prescriptions in \citet[eqs. (1) and (3)][]{cardelli1989}
to get $A_\lambda/A_V$ where $A_\lambda$ is the extinction in a
passband centered on wavelength $\lambda$. For the \gaia~G band,
$\lambda = 637$~nm; the \gaia\ Blue Photometer (BP) and 
Red Photometer (RP) bands are centered on 532~nm and 797~nm, 
respectively \citep{jordi2010}. The result is
\begin{equation}
E(\brcolor) \approx 1.355 E(B-V) \ \ \ \text{and} \ \ \ A_G \approx
0.848 A_V.
\end{equation}
Since \citet{schlegel1998} and \citet{schlafly2011} provide total dust
screening in the plane of the sky averaged over degree scales, $A_G$
and $E(\brcolor)$ for nearby sources at low Galactic latitudes are
approximate upper limits.

After correction for reddening, all but one of these fast stars has
color and luminosity indicative of the low metallicity, low
surface-gravity late-type giants identified by \citet{hattori2018b}
and \citet[see also \citealt{du2018}]{hawkins2018}. Even with
significant adjustments to our reddening correction, our interpretation
of these sources as late-type giants remains the same.

Of the fast stars, \gid{5932173855446728064} is the most reddened. It
lies within a few degrees of the Galactic plane and is the closest of
the fast stars to the Sun \citep[$\dsun=2.2$~kpc; see
  also][]{marchetti2018b}.  The 2-D V band extinction from
\citet{schlafly2011} at this source's location is $E(B-V)=0.88$, which
is likely an upper bound. A refined estimate, based on the 3-D,
low-latitude map by \citet[which we access through the
  \citealt{bovy2015} \texttt{mwdust} package]{marshall2006} gives
$E(B-V)=0.43$. Since small-scale variations in the dust
\citep[e.g.,][]{minniti2018} may impact the actual reddening of this
source, we indicate both estimates in Figure~\ref{fig:gcvelhr} (the
\citet{schlafly2011} and \cite{marshall2006} reddening agree to within
0.05 mag for other high-speed, low-latitude sources reported
here). Despite these uncertainties, the corrected $M_G$ and $\brcolor$
color are consistent with an early-type star either on the main
sequence or one that is just evolving off it. Follow-up spectroscopy
would distinguish these possibilities.

\begin{figure}
%\centerline{\includegraphics[width=4.5in]{gcvelverr.pdf}}
\centerline{\includegraphics[width=5in]{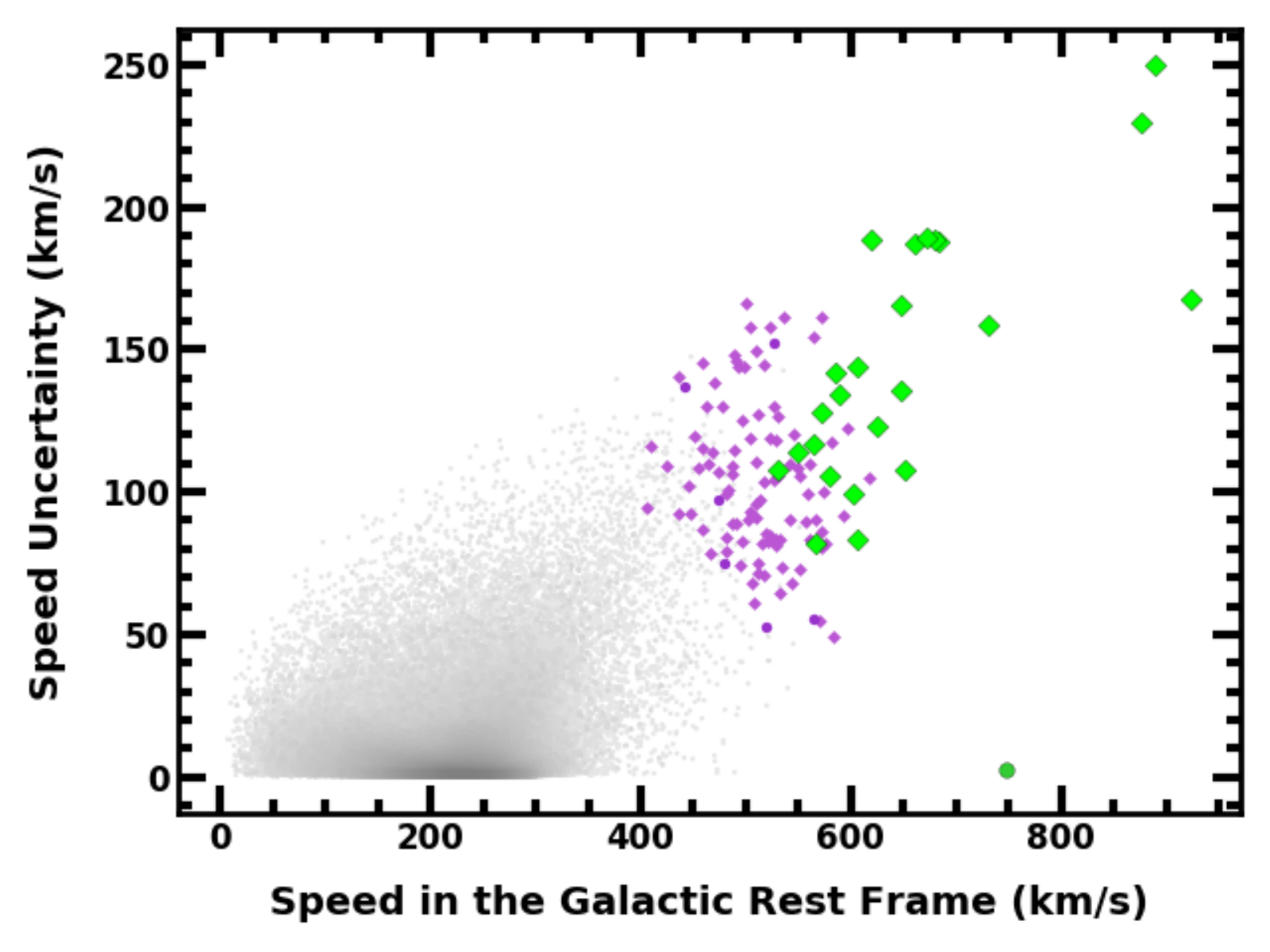}}
\caption{\label{fig:gcvelverr} The uncertainty in Galactic rest-frame
  speed as a function of distance from the Galactic Center. The
  meaning of the symbols is as in Figure~\ref{fig:gcvelhr}: Lime
  symbols are unbound candidates, magenta symbols are marginally
  bound, and the gray shading indicates the number density of the bulk
  of the $\sim$1.5M sources in our 6-D sample.}
\end{figure}

In Figure~\ref{fig:gcvelhr}, the fastest stars are among the most
luminous objects with \qual\ parallaxes and complete velocity
measurements.  Compared to the bulk of the $\sim 1.5$M stars in
Figure~\ref{fig:gcvelhr}, our high-speed candidates typically have a
large heliocentric distance.  All but one lie outside of $\dsun =
5$~kpc.  Because it is more challenging to acquire high quality
astrometry for more distant objects, it is important to consider
whether these high speed stars have measurement uncertainties similar
to those of more slowly moving stars.  Figure~\ref{fig:gcvelverr}
shows that our fastest-moving stars, with one exception, have
uncertainties in Galactocentric speed that scale with the speed. The
key message from this figure, along with the color-magnitude data, is
that our fastest moving objects are the most distant, most luminous
ones with the biggest uncertainty in speed.

%\newpage
\subsection{Orbital characteristics}

\begin{figure}
%\centerline{\includegraphics[width=6.5in]{gcvel2x2.pdf}}
\centerline{\includegraphics[width=6.5in]{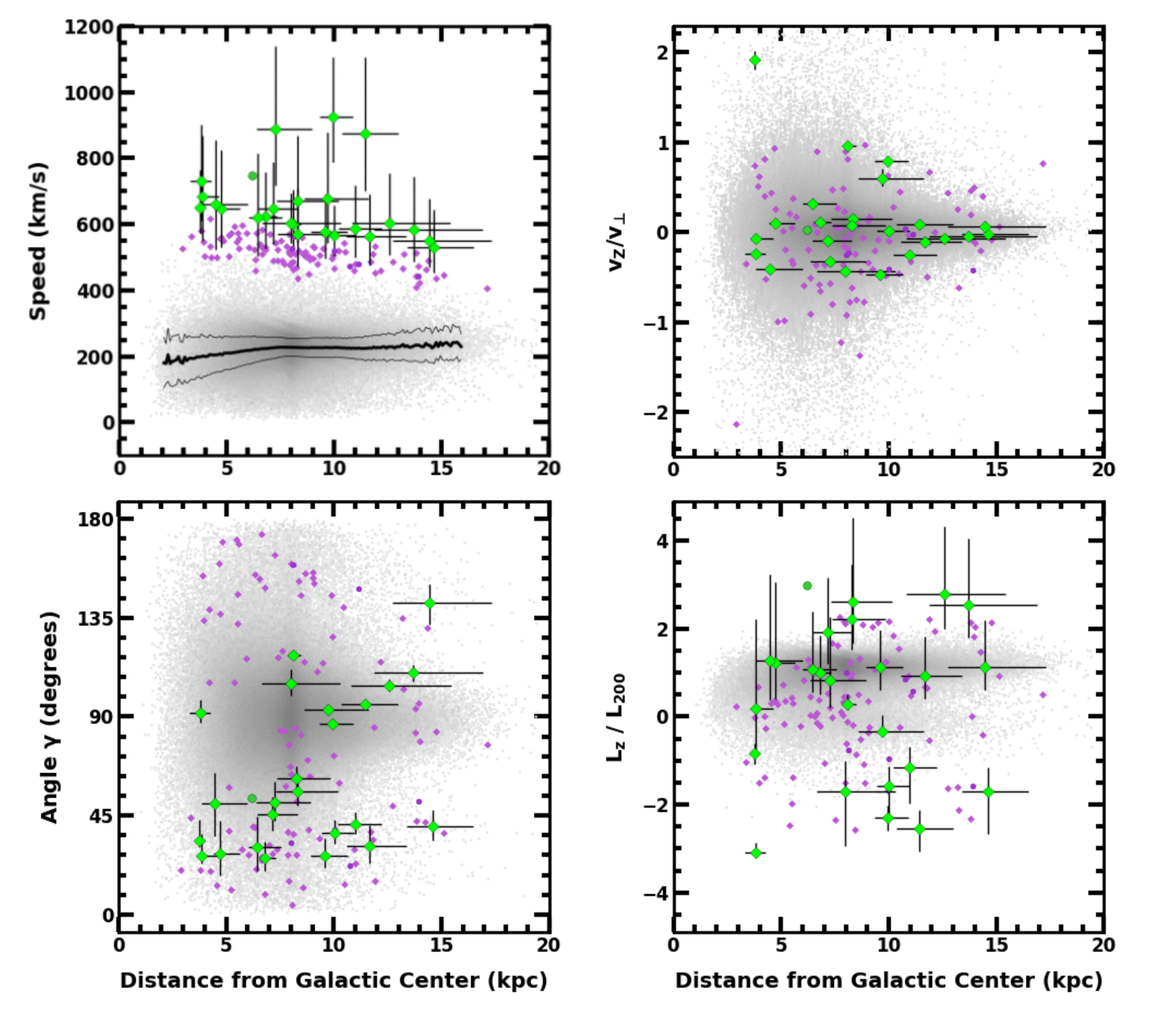}}
\caption{\label{fig:gcvelgc} Orbit characteristics as a function of
  distance from the Galactic center of sample stars. As in
  Figure~\ref{fig:gcvelhr}, lime-colored points represent unbound
  candidates ($\pub\ge 0.5$); errorbars mark the 16-th to 84-th
  percentiles in the position and velocity distributions. The magenta
  points are nominally bound stars with at least 20\% probability of
  being unbound ($0.2\le \pub <0.5$).  The circle points show sources
  with the \dupsrc\ flag set, whereas diamonds indicate stars for
  which this flag is not raised.  The light gray points are the
  remaining stars in our 6-D sample. The upper left panel shows
  Galactocentric speed, and the lower left shows the direction of
  travel relative to an outbound radial orbit.  The upper right plot
  is the vertical speed relative to the plane-of-the-disk speed, and
  the lower right plot shows the $Z$-component of the angular momentum
  relative to the value characteristic of circular orbits in the disk
  ($L_200$).  Together, these plots build a picture of the high-speed
  stars as mostly outliers of the halo and disk populations.  As a
  group, these stars also tend to be outgoing ($\gamma < 90^\circ$ in
  the lower left panel).  Their vertical motion is typically less than
  in the plane of the disk (upper right panel), and they tend to
  corotate with the Galactic disk (lower right panel).  Despite
  significant uncertainties, the fastest objects are distinct from the
  bulk of sample, which has a mean speed near 200~\kms (thick line in
  the upper left plot) and 1-$\sigma$ range of $\lesssim 100$~kms
  (thinner lines).  }
\end{figure}

To illustrate how the orbits of the fastest stars compare with the
bulk of the $\sim$1.5M stars in the sample, we present a few different
views of the data in Figure~\ref{fig:gcvelgc}. We look at the speed of
stars and their direction of travel in ways that highlight the
difference between disk stars, the halo population, and unbound HVS or
HRS orbits.  In the Figure, we distinguish between the 25 unbound
candidates (lime-colored symbols), the 101 nominally unbound objects
(magenta circles), and the roughly $1.5$~million remaining stars.

The Galactic rest-frame speed of stars as a function of distance from
the Galactic Center (Figure~\ref{fig:gcvelgc}, upper left panel) gives
an overall perspective of bound versus unbound candidates. The bulk of
stars in the sample have considerably slower speeds, around 200~\kms.
The dispersion is also low, $\sim 50$~\kms\ in the Solar
neighborhood, suggesting that most of these stars are part of the
disk. The dispersion creeps upward at larger distances and toward the
Galactic Center at least in part because the uncertainty in speed
($\vgcerr$) tends to increase with distance from the Sun.  

Figure~\ref{fig:gcvelgc} also provides information about the
orientation of stellar orbits. The lower left panel shows $\gamma$,
the angle of travel relative to a purely radial trajectory, a measure
tuned to select high-speed objects that originate from the Galactic
Center ($\gamma = 0^\circ$). The plot of $\vzvperp$ (upper right
panel) measures the degree to which orbits are confined to the disk
plane; $\LZ/L_{200}$, the $Z$-component of angular momentum relative
to typical disk star values (lower right panel), provides information
about the sense of orbital rotation compared with the disk. Together
these panels illustrate that the majority of the $\Nsamp \approx 1.5$M
stars shown in Figure~\ref{fig:gcvelgc} are associated with the disk:
stars cluster around $\gamma = 90^\circ$, $\vzvperp = 0$ and
$\LZ/L_{200} = 1$.  Another distinct population of stars is more
evenly distributed in $\gamma$ and $\vzvperp$ and has values of
$\LZ/L_{200}$ centered around zero. These are likely halo stars.

Within Figure~\ref{fig:gcvelgc}, the marginally bound stars (magenta
symbols) generally track our expectation of halo stars.  There are
roughly as many incoming as outgoing objects, with $\gamma$ values
loosely clustered about $\sim 30^\circ$ and $\sim 150^\circ$. The mean
and dispersion of $\vzvperp$ is characteristic of isotropic orbits.
Furthermore, comparable numbers of these stars corotate with the
Milky Way's disk as counterrotate.

Among the 25 unbound candidates (lime-green symbols), the distribution 
suggests a trend toward orbits that are outgoing from the Galactic Center 
and co-rotating with the disk. Roughly two-thirds are on outgoing orbits 
(17 of 25 stars); a majority corotates with the disk (16 stars). Almost 
half of these sources have both of these characteristics (12 objects). 
Thus, while no compelling HVS candidates emerge, there is a hint that of 
some of the unbound candidates are HRSs.

\subsection{Are the outliers really unbound?}\label{subsec:really}

The top 25 high-speed stars may include unbound HVS and HRS candidates
\citep{kenyon2014, kenyon2018}.  However, they may simply be 
slower-moving bound objects with large errors, drawn from the large 
pool of almost 1.5 million stars. When the unbound probability of a 
source has a modest value, different from unity (e.g., $\pub\sim 0.5$), 
we cannot distinguish the unbound cases from the bound outliers. Even 
when $\pub$ for a candidate is close to unity, we must consider whether
it is a rare outlier of the bound population.

To develop an approach for distinguishing a truly unbound object from
a bound outlier, we examine each unbound candidate and ask whether
known bound stars could have been measured with an unbound probability
as large as the candidate's.  Thus, we run through the long list of
nominally bound stars ($\pub < 0.5)$, identifying those with
uncertainties in Galactic distance and speed that are similar or
better than the candidate's.  Then, by sampling the error
distributions of these bound objects, we determine the probability,
$\pbo$, that each bound star would be observed to be an unbound object
like the candidate.  If the candidate were just an outlier, then we
expect $\pbo$ to be significant for at least some bound stars; if no
bound star has any likelihood of being observed as an outlier with
similar properties ($\pbo=0$), then the candidate stands out as a true
unbound star.

In our algorithm for quantifying whether a star is unbound or an
outlier, we work with the list of bound stars that have the same or
smaller relative uncertainties, $\rgcerr/\rgc$ and $\vgcerr/\vgc$,
compared to the candidate. For the $i$-th bound star on the list,
Quasi-Monte Carlo trials give samples of the state vector in Galactic
coordinates, based on the measured astrometry and uncertainties. A
separate Quasi-Monte Carlo estimate gives the unbound probability
$\pub$ for each state vector. The fraction of trials that give $\pub$
equal to or greater than that of the unbound candidate is our estimate
of $\pbo$ the chance that a bound star would be identified as unbound,
like the candidate.

The likelihood that an unbound star is not just an outlier comes from
tallying up the possibilities that individual bound stars might be
perceived as unbound,
\begin{equation}
\pbx = 1 - \prod_i (1-\pbo^{(i)}),
\end{equation}
where the index $i$ runs over the list of bound stars.
Similarly, the typical number of bound stars that are expected to
be outliers like the candidate is
\begin{equation}
\Nbx \sim \sum_i \pbo^{(i)}.
\end{equation}
Thus, if none of the bound stars has as much of a chance of being an 
outlier as the unbound candidate ($\pbo=0$ for all $\Nsamp\approx$1.5M stars),
$\Nbx = 0$. The candidate then has zero probability of being a bound 
outlier; it is probably unbound.  If $\pbx$ is much less than unity, 
then we cannot distinguish the candidate from the unbound population.

%To perform this calculation for each unbound candidate, we generate
%4,096 quasi-random samples of Galactocentric distance $\rgc$ and speed
%$\vgc$ for all of the nominally bound stars, based on the measurement
%uncertainties (as in \S\ref{sec:sampsel}). 
Applied to our high-speed sample, this analysis yields two sources,
\gid{5932173855446728064} and \gidinlist{1383279090527227264}, as the
most likely unbound stars.  For the first source, the likelihood of
drawing bound outliers with its orbital characteristics and
measurement errors is formally zero ($\pbx = 0$).  None of the roughly
6~billion Quasi-Monte Carlo samples yields as extreme an outlier
($\Nbx = 0$), primarily because of the candidate's high speed with
small uncertainty ($747\pm 3$~\kms). The second source,
\gid{1383279090527227264}, also has a formal 100\%\ probability of being an
unbound star and not an outlier of the bound population.

Of the remaining sources, \gid{1478837543019912064} has the greatest
chance of being an unbound star. There is an 86\% chance that it is
not just a bound outlier, although we expect that one star from a
bound population would be an outlier as extreme as this star ($\Nbx =
0.99$).  \gid{6456587609813249536} is a runner-up, with a 49\%\ chance
of being unbound given the pool of other sources ($\Nbx = 0.98$). The
remaining 21 objects have more than a 99\% chance of being bound
outliers.  These preliminary results do not indicate that these
candidates are really bound; they just cannot be distinguished from
bound outliers \citep[see also][]{bromley2006,brown2007a,kenyon2018}.

We summarize aspects of the two main candidates, including their
orbits in the Galaxy.  We use a simple, fixed-timestep, fourth-order orbit
integrator to estimate how an object moves within our choice of
Galactic potential \citep[the integrator is similar to the one
  described in][]{kenyon2018}.

\begin{itemize}
\item
\gid{5932173855446728064}: This star has a high radial velocity with
low uncertainty ($\rv = 747\pm 3$~\kms)
\citep[cf.][]{marchetti2018b}. At a Galactocentric distance of $\rgc =
6.2\pm 0.1$~kpc, its radial motion alone indicates that it is
gravitationally unbound. With position $(X,Y,Z)= (-6.1,-1.1,-0.1)$~kpc
and speed $(U,V,W)=(-549,506,18)$~\kms, it is moving outward and with
the rotation of the Galaxy, skimming the underside of the disk. A
traceback of its orbit gives a closest approach to the Galactic Center
of 4.8~kpc, at a distance of about 0.2~kpc below the plane.  Because
the star's orbit is deflected upward by the gravity of the Galaxy, the
star formally crossed the disk on the far side from the Sun ($\rgc
\sim 20~kpc$), within the past 50~Myr.  The reddening and extinction
also suggest that this source may be an A-type main sequence star.
Thus it may well be a hyper-runaway star of Galactic origin. However,
this source is in a crowded field and has been flagged as a
\dupsrc\ in the \dr2\ archive.  Follow-up observations would resolve
these ambiguities.

\item
\gid{1383279090527227264}: At a distance of $10.0\pm 0.9$~kpc and with
a speed of $924\pm 168$~\kms\ relative to the Galactic Center, this
object is likely to be unbound \citep[see also][]{marchetti2018b}. It
is situated well above the disk ($X,Y,Z = -5.7,5.1,6.4$~kpc) and its
orbit arches upward and against the rotation of the disk
($U,V,W=-91,-719,573$~\kms).  An orbit calculation back to the disk
suggests an intersection at a distance just inside of 14~kpc from the
GC about 15~Myr ago. An integration farther back in time gives an
orbit that crosses within about 15~kpc of the Large Megellanic Cloud
roughly 70~Myr ago.  (One other star, \gid{6492391900301222656}, has a
similar closest approach to the LMC.)  Thus, this star has promise as
a hyper-runaway star candidate, or even an escapee from the
LMC. Despite these intriguing possibilities, the kinematics, compared
with the bulk of the 6-D samplei, and the source's position in the
color-magnitude diagram remind that it is probably a statistical
high-speed outlier of the halo's late-type giant population.
%consistent with our analysis ($\pbx =
%0.49$).

\end{itemize}

\subsection{The impact of distance estimation}

The results presented in this section come from a Quasi-Monte Carlo
analysis in which heliocentric distances and parallaxes are related
simply by $\dsun = 1/\pax$. A Bayesian approach (\S\ref{sec:sampsel})
applied to the \qual\ sample alters Galactocentric position and
velocities only modestly, but the impact on the unbound probabilities 
$\pub$ can be more substantial. For example, a prior for constant
source density within roughly 20~kpc of the Sun ($L = 40$~kpc in
equation~(\ref{eq:prior}), motivated by HVS ejection models
\citep{bromley2006}), yields 99 stars with $\pub\ge 0.5$, including all
of the sources listed in Tables~1 and 2. This increase arises because
the prior causes distance estimates to increase, in accordance with
the assumption that there are more sources at larger distances, at
least for the parallax range of the \qual\ stars.  This change raises
the tangential speed $\vtpm$ and the unbound probability $\pub$.

A Bayesian prior that roughly follows the distance distribution of the
bulk of the \gaia\ stars (equation~(\ref{eq:prior}) with $L =
1.5$~kpc; cf.\ \citealt{bailer-jones2018}) tends to shift raw
parallaxes toward the peak of the distance distribution of the bulk
($\dsun = 2L$). More distant stars are estimated to be closer to the
Sun, decreasing the tangential speed and the unbound probability The
opposite happens for close-in stars. Both effects are present in our
data. Analysis with this prior yields 16 stars with $\pub \ge 0.5$; Of
the 25 stars in Table~2, 13 appear in the new catalog. The drop-outs
are have $\pub < 0.75$, and are at median distances $\dsun >
8$~kpc. Three new candidates emerge (\gid{3905884598043829504},
\gid{3705761936916676864} and \gid{6516009306987094016}), all with
comparatively small distances, below about 5~kpc, and unbound
probabilities near our 50\% threshold.

Adopting a larger value of scale length $L$ in the exponential prior,
as in \citet{marchetti2018b}, gives results that are more similar to the
ones presented here. The heliocentric distance estimates between the
two cases change typically by a few percent or less. The exponential
model draws in 11 new objects with $\pub$ near the 50\%\ threshold,
and of the 25 high-speed stars identified here, only one star, the
object with the lowest $\pub$, falls below the threshold as a result
of the prior.

All of the distance estimation methods explored here yield the same
set of stars with high probability of being unbound to the Galaxy,
with $\pub > 0.75$ in Table~\ref{table:hvsgc}. Thus, the
identification of the most promising candidates is insensitive to the
details of distance estimation, which is our motivation for selecting
the \qual\ sample.

\section{INFERRING 3-D VELOCITY FROM PROPER MOTION OR RADIAL VELOCITY}
\label{sec:getvel}

\begin{figure}[htb]
%\centerline{\includegraphics[width=4.5in]{gcvelrv.pdf}}
\centerline{\includegraphics[width=5in]{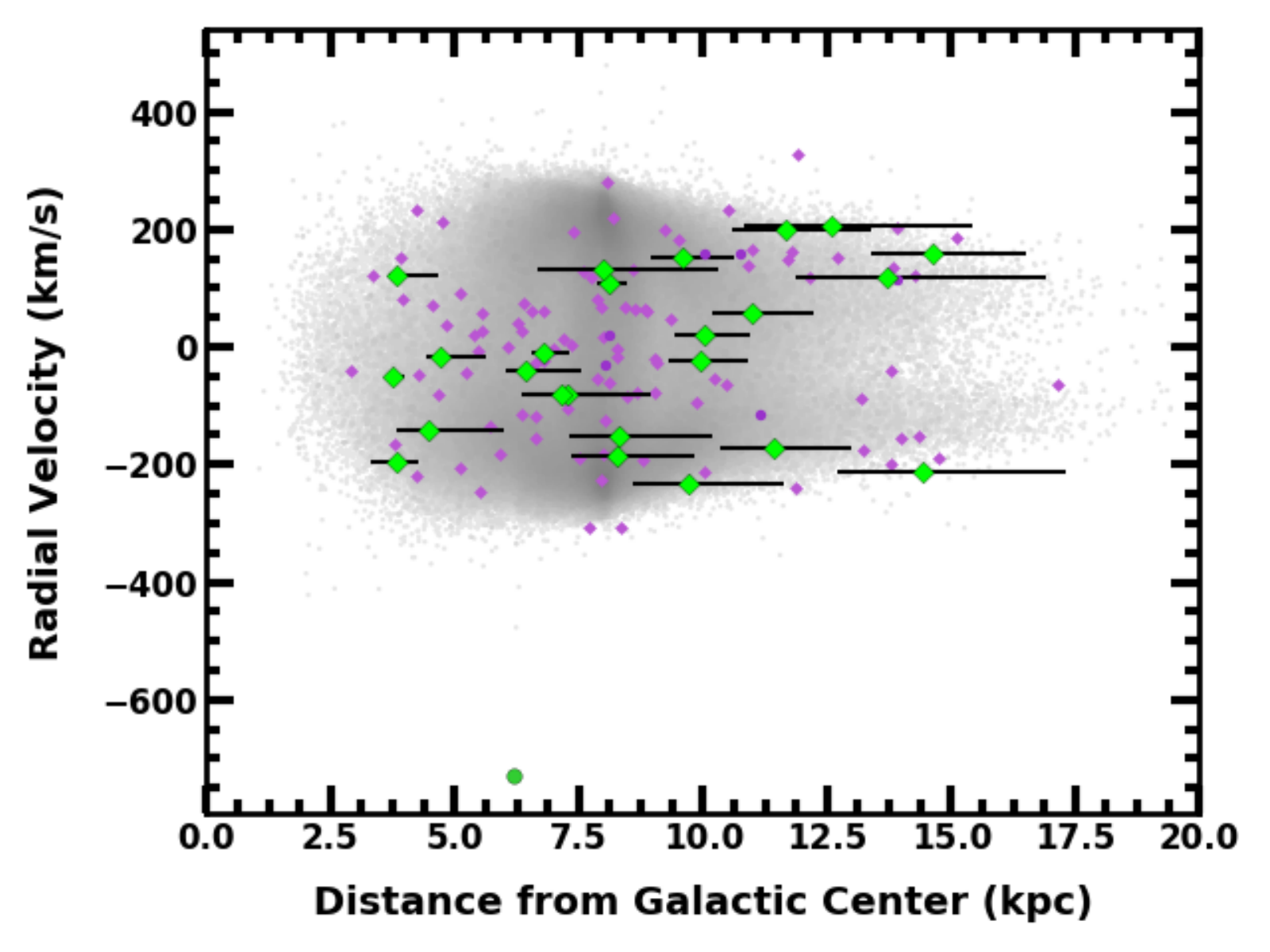}}
\caption{\label{fig:gcvelrv} The radial velocity of stars, corrected
  for Solar motion. The symbols are the same as in the preceding
  Figures; lime-green points are unbound ($\pub\ge 0.5$), magenta
  points are marginally bound ($0.2\ge\pub<0.5$), and the remaining
  1.5M stars are represented in gray. Except for two outliers, the
  unbound or marginally bound stars are mixed in with the
  slower-moving objects. Radial velocity is not a strong discriminant
  for high-speed stars in this sample.  }
\end{figure}

In addition to finding unbound candidates, our 6-D sample allows
evaluation of other measures of velocity as indicators of stellar orbits.  
Figure \ref{fig:gcvelrv} shows the radial velocity in the Galactic frame, 
$\vrrv$ ($\rv$ corrected for Solar motion). One star, the top 
unbound candidate \gid{5932173855446728064}, stands out with a high radial 
speed, $\vrrv > \vesc$. With ($\ell,b)=(329.94^\circ,-2.70^\circ)$, it lies
inside the Solar circle, roughly in the direction of the Galactic Center 
relative to the Sun, and is approaching the Sun with a closing speed of over 
600~\kms\ \citep{marchetti2018b}.

Figure \ref{fig:gcvelrv} reveals a second outlier,
\gid{1364548016594914560}, a marginally bound star (the uppermost
magenta point in the Figure). This source lies beyond the Solar circle
and is on an orbit that is nearly radially outward from the Galactic
Center. Although it was tagged as a hypervelocity candidate by
\citet{marchetti2018b}, it is not included in our top-25 list; 
our choice of Galactic potential suggests that this star is most
likely bound to the Galaxy \citep[see also][]{brown2007a}.

Other than the two outliers, the stars shown in Figure
\ref{fig:gcvelrv} have radial speeds of less than $\sim$300~\kms.  Thus,
the $\vrrv$ values of the high-speed outliers give little indication of
unusual motion.  Radial velocities alone do not consistently
distinguish high-speed stars from bound disk or halo stars at the
modest Galactocentric distances discussed here.

\begin{figure}[htb]
\centerline{\includegraphics[width=5in]{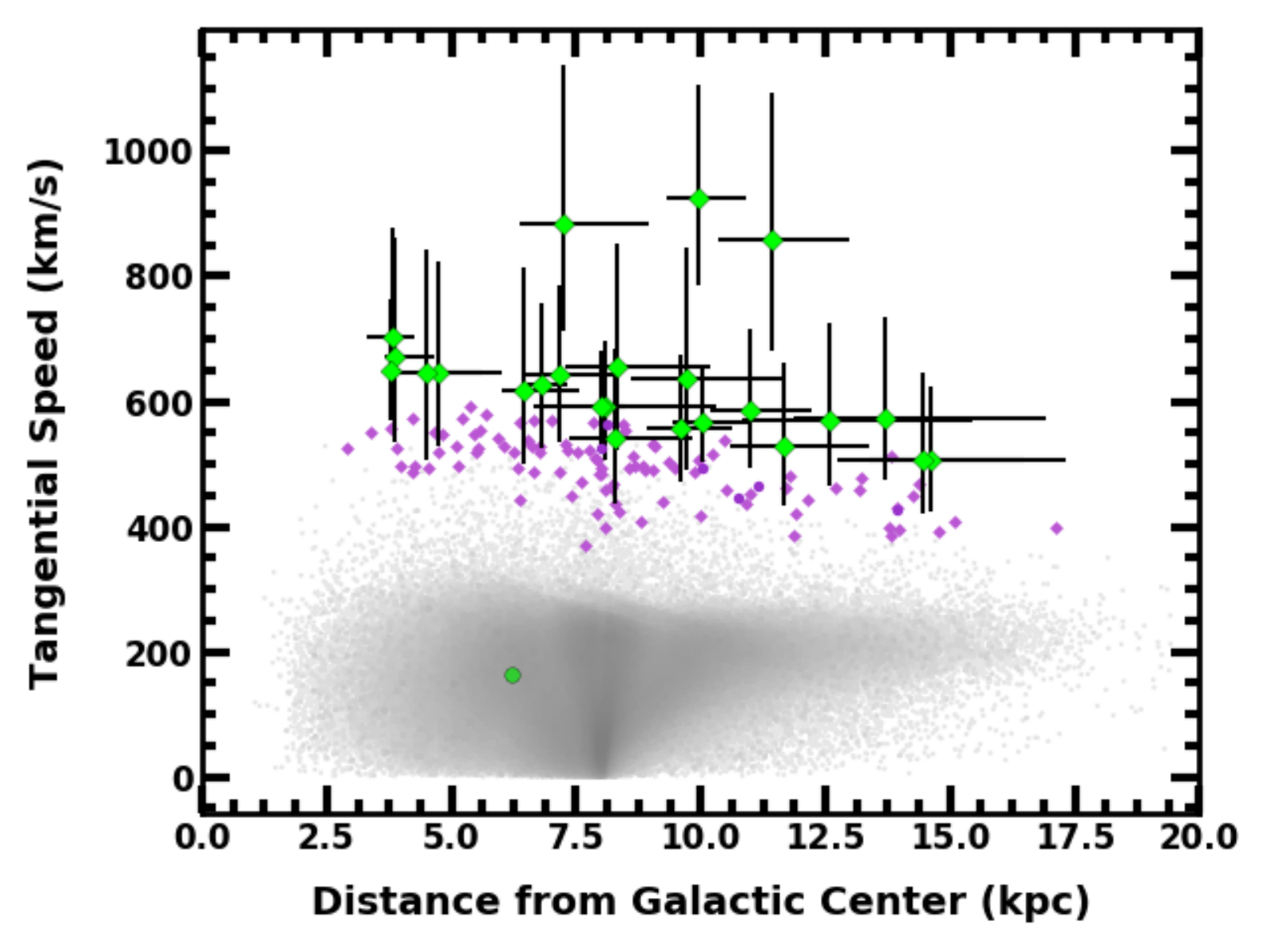}}
\caption{\label{fig:gcvelpm} The tangential speed of stars, corrected
  for Solar motion.  The symbols are the same as in the previous plot.
  Here, the tangential motion leads to a clustering of high-speed
  stars roughly sorted by probability of being unbound to the Galaxy. 
  The one exception is the lime-colored point below $\vtpm =
  200$~\kms; its motion, relative to the Sun, is predominantly radial.
}
\end{figure}

Figure \ref{fig:gcvelpm} shows the tangential speed of stars in our
sample, $\vtpm$, derived from parallax and proper motion, and
corrected for Solar motion. Nearly all of the high-speed stars that
we identify as either unbound or marginally bound lie in the upper
envelop of the speed distribution.  Thus, in contrast to radial
velocity, proper motion leads to a more robust indicator of speed
relative to the Galactic Standard of Rest, at least for stars within
$\sim$10~kpc of the Sun.  The one exception to this rule is the
radial-velocity outlier, \gid{5932173855446728064}, which exhibits
comparatively little tangential motion.

\begin{figure}[htb]
\centerline{\includegraphics[width=5in]{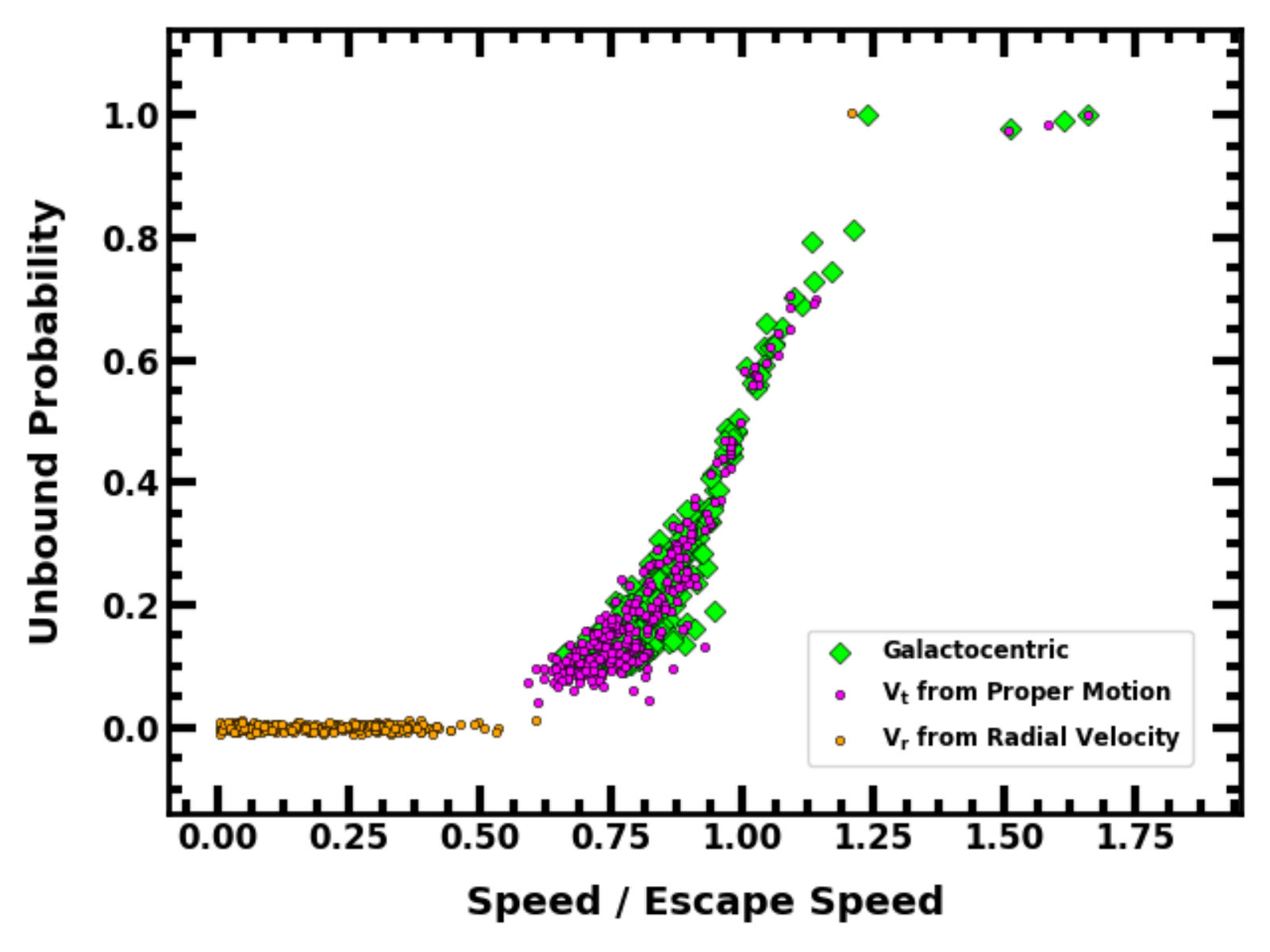}}
\caption{\label{fig:gcvelvescpub} Probability that a star is unbound 
  to the Galaxy, as a function of speed in the Galactic rest frame. 
  The speed of each star is derived in three ways: from radial 
  velocity $\vrrv$ alone (orange triangles), tangential speed $\vtpm$ 
  alone (magenta circles), and the total Galactocentric speed $\vgc$ 
  (lime diamonds; each star has three points associated with it).  
  Only the last set of points, derived from both radial and tangential 
  motion, gives the true unbound probability, $\pub$.  The tangential 
  data track the true probability reasonably well. The radial velocity 
  data fail miserably, but not completely, in predicting $\pub$. The 
  one exception is the high-speed radial velocity outlier (the upper 
  orange triangle, which corresponds to the green diamond to its 
  immediate right). Only stars with $\pub > 0.01$ are considered in 
  this analysis.
}
\end{figure}

For stars at the modest heliocentric distances in our sample,
tangential velocity is a good measure of the probability that a star
is unbound. This result contrasts with more distant samples, where the
radial velocity selects robust samples of unbound stars
\citep[e.g.,][]{brown2006a, brown2009b, brown2014}.
Figure~\ref{fig:gcvelvescpub} shows the Galactocentric speed of stars
in our sample as a function of unbound probability $\pub$.

We also show the unbound probability as determined from radial and
tangential speeds separately. The distributions of these sets of
points illustrate that tangential speed is a reasonable indicator of
$\pub$. Radial velocity measurements on their own mostly fail to
predict $\pub$. Thus, for nearby stars with distances of 10--15~kpc,
selection according to proper motion and parallax is an efficient way
to find high-speed stars.  In our sample, identifying stars with
tangential speeds exceeding 0.75 times local escape speed selects
92\%\ of the stars that are unbound to the Galaxy with $\pub$ = 20\%
or higher, with a false detection rate of 43\%. In contrast,
identifying nearby, high speed stars solely by radial velocity yields
only one candidate (\gid{5932173855446728064}).

\section{NEW CANDIDATES IN 5-D}\label{sec:newvtpm}

From the 6-D analysis of \S\ref{sec:getvel} and the calculations of
HVSs in \citet{kenyon2018}, measurements of tangential velocity
provide a good indicator of nearby (10--15~kpc) high-speed stars,
including those on orbits that emanate radially outward from the
Galactic Center.  We now consider stars in the full \dr2\ archive that
satisfy the astrometric criteria listed in \S\ref{sec:sampsel} yet do
not have radial velocity measurements. Our analysis of these sources
is as described in \S\ref{sec:sampsel}, except with no radial velocity
component.  Because we are most interested in fast stars, we consider
a subset of these sources that have an observed tangential speed,
corrected for Solar motion, of $\vtpm > 400$~\kms\ (approximately 70\%
of the escape speed in the Solar neighborhood). This sample of 9939
fast-moving objects contains the most promising unbound candidates in
5-D.

\begin{figure}[htb]
%\centerline{\includegraphics[width=4.5in]{gcvelhrx.pdf}}
\centerline{\includegraphics[width=5in]{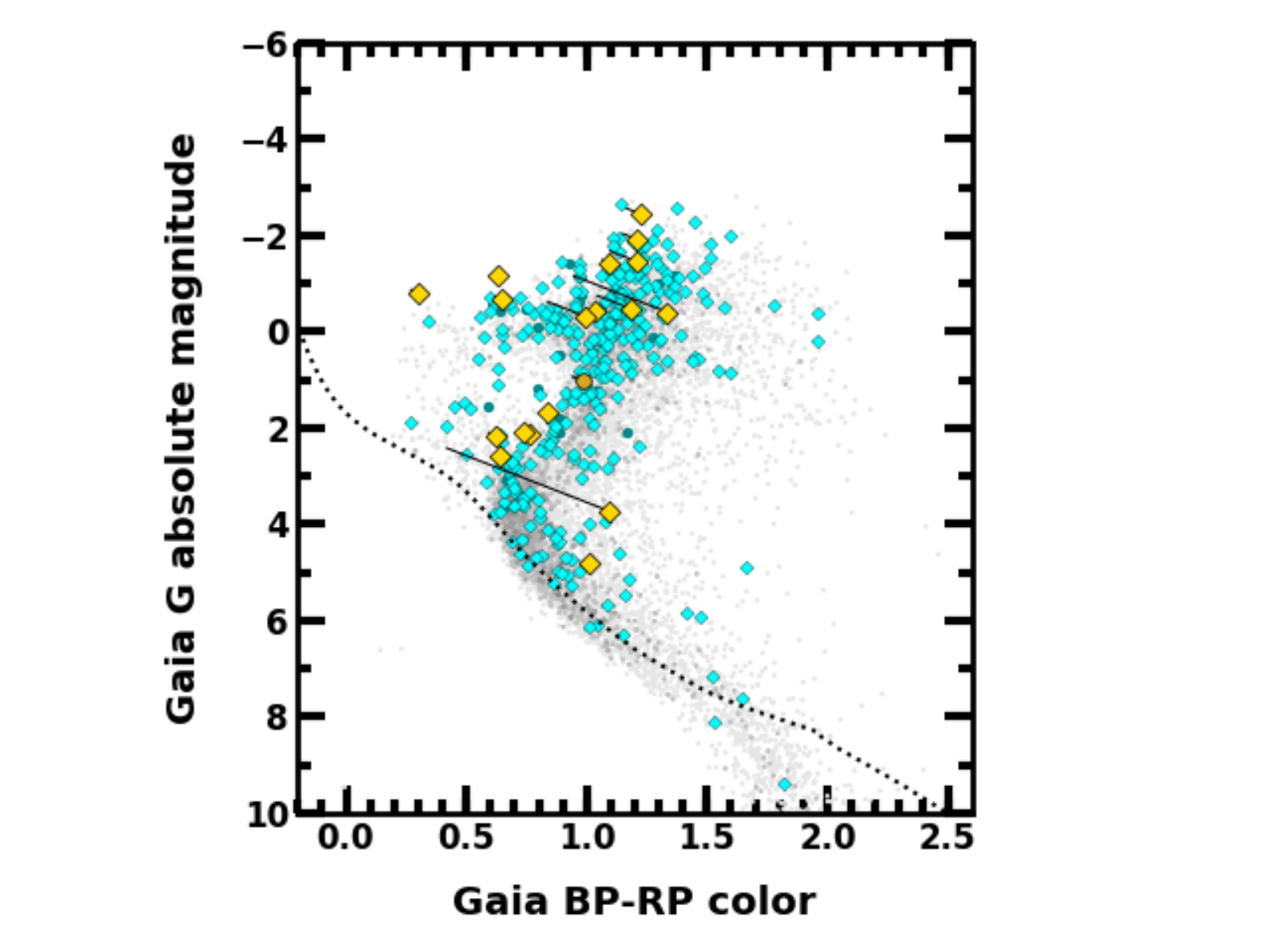}}
\caption{\label{fig:gcvelhrx} The color-magnitude diagram of stars in
  the \dr2\ with \qual\ astrometry but no radial velocity measurement.
  The dotted line shows the position of the main sequence.  All stars
  with $\vtpm > 400$~\kms\ are shown in gray. The cyan diamonds
  indicate stars with $\pub\ge 0.5$; the yellow-gold diamonds plot the
  subset with $\pub \ge 0.95$. Darker-shaded circles show sources that
  have the \dupsrc\ flag set.  Thin solid lines
  indicate reddening vectors. Unlike the 6-D data, the fastest stars
  include those near the main sequence as well as late-type giants.}
\end{figure}

\begin{figure}[htb]
%\centerline{\includegraphics[width=4.5in]{gcvelpmx.pdf}}
\centerline{\includegraphics[width=5in]{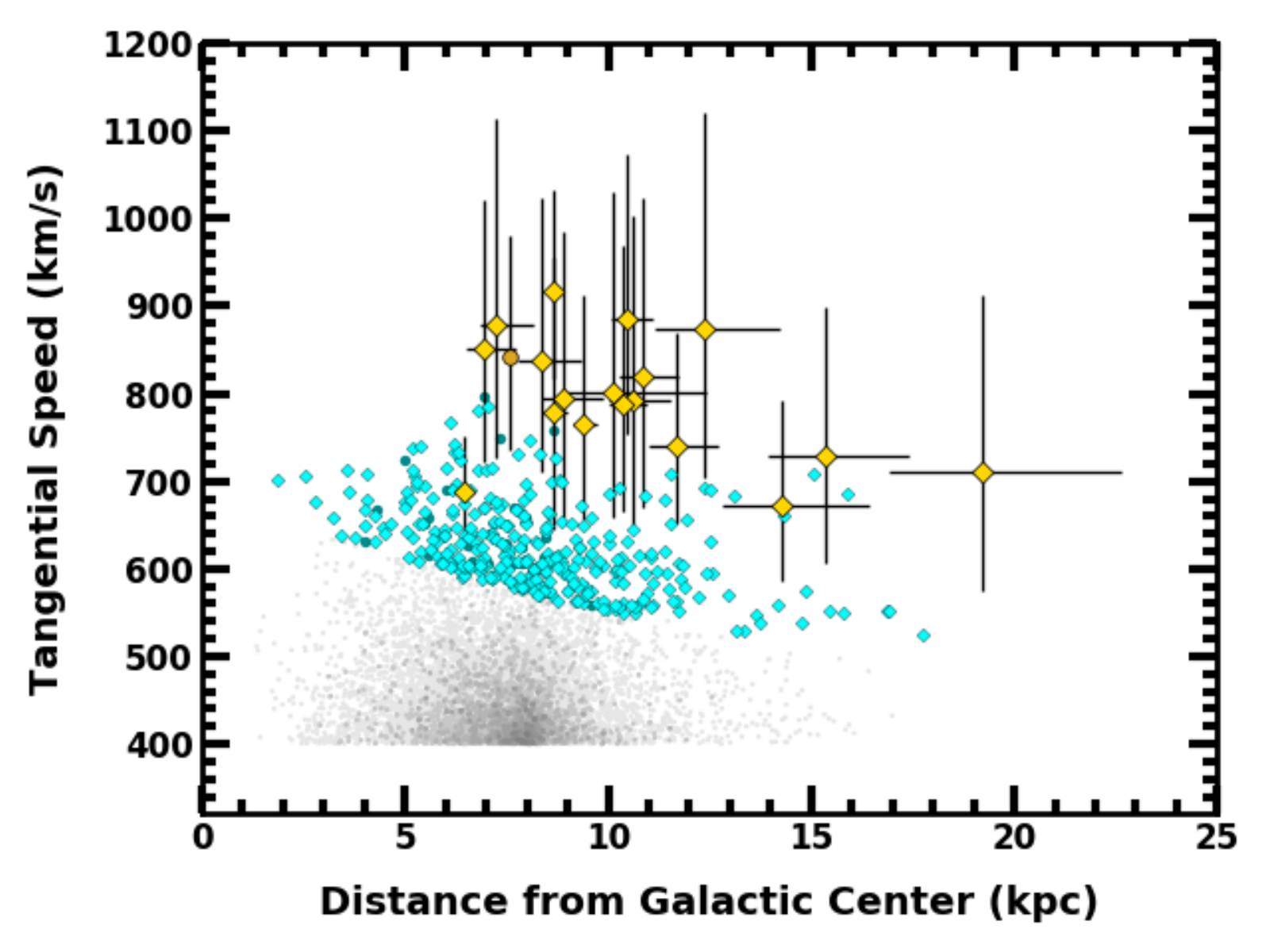}}
\caption{\label{fig:gcvelpmx} The Galactic-frame tangential speed of
  fast-moving stars in the \dr2\ with \qual\ astrometry but no radial
  velocity measurement.  As in Figure~\ref{fig:gcvelhrx}, yellow-gold
  symbols plot stars with probability $\pub\ge 0.95$; cyan points 
  correspond to stars with $0.5 \le \pub \le 0.95$; gray points indicate 
  stars with $\vtpm > 400$~kms and $\pub \le 0.5$. This sample and 
  the theoretical analysis of \citet{kenyon2018} suggest that the 
  high-speed stars are excellent candidates for HVSs and HRSs.}
\end{figure}

The proper motion sample includes 343 stars with $\ge$
50\%\ likelihood of being unbound ($\pub\ge 0.5)$. Of these, 19 are
unbound with 95\% confidence on the sole basis of tangential velocity
and astrometric uncertainties.  Table~\ref{table:hvsvt} summarizes the
properties of these 19 fastest objects. Figure~\ref{fig:gcvelhrx}
provides brightness and color information of all stars in this sample,
with indicators of reddening and extinction from the IPAC/IRSA DUST
service. The most significantly reddened sources
(\gid{2946665465655257472} and \gid{1820299950021811072}) are both
within 10$^\circ$ of the Galactic plane; nonetheless the values of
$E(B-V)$ provided by IRSA (0.497 and 0.287, respectively) are within
0.1~mag of the estimate from the 3-D \texttt{Combine15} maps in the
\texttt{mwdust} package \citep{bovy2016}.  Figure \ref{fig:gcvelpmx}
shows the Galactocentric distance and speed information for all stars
in this sample.

From this subset of the \dr2\ archive, we conclude that the high-speed 
stars in 5-D are more broadly representative of the general population 
of stars in the Milky Way. We have not evidently selected an exclusive
set of HVS or HRS candidates. Instead, we speculate that these stars, 
like the majority of our high-speed candidates in 6-D, are outliers of 
a bound distribution. Radial velocity measurements of these top candidates 
are required to test this interpretation.

\begin{deluxetable}{lrrcrrrc}
\tabletypesize{\scriptsize}
\tablecaption{High-speed stars: tangential speed selection.
\label{table:hvsvt}}
%\tablewidth{5.5in}
\tablehead{
\colhead{\gaia\ DR2} & 
\colhead{G} & \colhead{$\brcolor$} &
\colhead{($\ell$,$b$)} &
\colhead{$\dsun$} & \colhead{$\rgc$} & \colhead{$\vtpm$} &  
\colhead{$\pub$}
\\ 
\colhead{designation} & 
\colhead{(mag)} & \colhead{(mag)} & 
\colhead{(deg)} &
\colhead{(kpc)} & \colhead{(kpc)} & \colhead{(\kms)} &
 \colhead{\ } 
}
\startdata
1540013339194597376 & 15.96 & 1.01 & (145.31, 68.25) &  1.70$\pm$0.16  &  8.7$\pm$0.1  & 917.2$\pm$107.1 & 1.00 \\
1240475894700468736 & 14.03 & 0.99 & ( 20.41, 67.83) &  3.99$\pm$0.47  &  7.6$\pm$0.1  & 842.6$\pm$124.1 & 1.00 \\
1570348658847157888 & 15.63 & 0.74 & (122.23, 62.07) &  5.07$\pm$0.79  & 10.5$\pm$0.6  & 885.7$\pm$179.0 & 1.00 \\
1312242152517660800 & 13.81 & 0.30 & ( 50.78, 40.14) &  8.39$\pm$1.52  &  8.3$\pm$0.9  & 837.0$\pm$172.1 & 0.99 \\
4727516205455716224 & 13.57 & 0.63 & (273.60,-51.39) &  8.84$\pm$1.26  & 11.7$\pm$0.9  & 740.7$\pm$117.2 & 0.99 \\
1586391907885793792 & 14.39 & 0.64 & ( 78.07, 57.23) & 10.34$\pm$2.32  & 12.3$\pm$1.9  & 873.4$\pm$244.0 & 0.99 \\
5778956291515661440 & 14.31 & 0.99 & (313.01,-19.17) &  8.45$\pm$1.39  &  6.9$\pm$0.8  & 852.1$\pm$168.5 & 0.99 \\
4798132614628163968 & 15.92 & 0.62 & (253.31,-35.03) &  5.69$\pm$1.24  & 10.9$\pm$0.9  & 819.0$\pm$194.7 & 0.99 \\
1820299950021811072 & 14.01 & 1.34 & ( 54.89, -6.08) &  7.66$\pm$1.57  &  7.3$\pm$0.9  & 877.3$\pm$208.9 & 0.99 \\
2946665465655257472 & 16.22 & 1.09 & (227.47, -9.16) &  3.17$\pm$0.66  & 10.4$\pm$0.6  & 786.7$\pm$173.6 & 0.98 \\
1527780516422382592 & 15.14 & 0.77 & (117.82, 74.69) &  4.03$\pm$0.58  &  9.4$\pm$0.3  & 765.5$\pm$135.7 & 0.98 \\
4535258625890434944 & 13.14 & 1.18 & ( 52.01, 14.00) &  5.31$\pm$0.41  &  6.4$\pm$0.0  & 688.4$\pm$55.2 & 0.98 \\
1702417150851952128 & 13.84 & 1.10 & (107.08, 37.42) & 11.39$\pm$2.19  & 15.4$\pm$1.9  & 728.7$\pm$160.8 & 0.97 \\
4783869234396531968 & 15.44 & 0.84 & (257.18,-38.70) &  5.71$\pm$1.15  & 10.6$\pm$0.9  & 792.1$\pm$200.6 & 0.97 \\
2154188852160448512 & 13.51 & 1.21 & ( 86.83, 24.49) & 12.23$\pm$2.33  & 14.3$\pm$2.0  & 671.9$\pm$110.6 & 0.97 \\
6524618551753693952 & 14.06 & 1.04 & (328.60,-64.49) &  7.86$\pm$1.50  &  8.9$\pm$1.0  & 795.0$\pm$190.1 & 0.96 \\
1484524973071001984 & 15.83 & 0.64 & ( 68.04, 68.64) &  4.50$\pm$0.77  &  8.6$\pm$0.3  & 779.1$\pm$170.1 & 0.96 \\
6358539652542070912 & 13.95 & 1.21 & (315.20,-37.99) & 12.08$\pm$2.72  & 10.1$\pm$2.2  & 800.7$\pm$217.1 & 0.95 \\
4805658359403594624 & 13.46 & 1.22 & (249.38,-32.52) & 15.25$\pm$3.61  & 19.2$\pm$3.3  & 710.0$\pm$194.2 & 0.95 \\
\enddata
\end{deluxetable}

The list in Table~\ref{table:hvsvt} with $\pub \ge 0.95$ changes in
a Bayesian analysis. As with the 6-D sample, a prior tuned for a
uniform source density within the Milky Way yields larger distance
estimates and is more permissive in the assessment of whether a star
is unbound. It yields 90 sources, including the 19 stars in the Table.
An analysis with an exponential prior and $L =1.5$~kpc
(equation~(\ref{eq:prior})) flags 17 stars with $\pub \ge 0.95$, 15
of which are in Table~\ref{table:hvsvt}. Two are new candidates
(\gid{6182941362050295424} and \gid{6698855754225352192}) within 6~kpc
of the Sun. The four from the Table that are missed by the Bayesian
analysis all have estimated median distances beyond 10~kpc.  All three
methods identify the same highest speed sources with $\pub > 0.98$.

\section{COMPARISON WITH THEORY}\label{sec:theory}

\citet[Figures 21 and 22 therein]{kenyon2018} predict that HVSs and 
HRSs are efficiently selected at heliocentric distances well beyond 
$\sim 10$~kpc with radial velocities. At these distances, tangential 
motion is not significant.  Closer to the Sun and to the Galactic 
Center, radial velocities may be significant, but will not typically 
differ much from the motion of bound halo stars. Tangential 
speeds, however, will be high for nearby HVSs and HRSs. Radial velocities 
select high-speed stars at large distances; proper motion selects the 
fastest-moving stars within $\sim 10$~kpc.

Our analysis, based on 6-D data but using only 5-D information for
comparison, confirms the geometrical argument in
\citet{kenyon2018}. All of the high-speed outliers, with one
exception, have large tangential speeds and nondescript radial
speeds. The exception, \gid{5932173855446728064}, is an unusual star,
seemingly coming nearly directly toward the Sun in the heliocentric
frame.  Figures~\ref{fig:gcvelrv} and \ref{fig:gcvelpm} illustrate
these effects.

We also have sought to identify HVS and HRS candidates in the
6-D data. Focusing on the top 101 high-speed stars, only
\gid{5932173855446728064} has a high probability of
being unbound, a low likelihood that it is an outlying bound star,
and spectral type suggesting a young main sequence star.
Other stars are more suggestive of sample outliers of the
halo's late-type giant population. None of the stars has 
a radial direction of travel ($\gamma$ near 0$^\circ$).
A Galactic Center origin for this population is not favored.

The small number of HVSs is expected. The sample region is small
(a radius of $\sim 10$~kpc), compared to the region of space explored
in HVS searches of the Milky Way halo \citep[out to $\sim
  100$~kpc;][]{brown2005, brown2006a, brown2007a, brown2014}. Theory
predicts few if any A- or B-type stars on hypervelocity trajectories 
in this small region \citep[e.g.,][]{hills1988, yu2003, bromley2006, 
kenyon2014, hamers2017}; observations support this assessment
\citep[e.g.,][]{brown2006a, koll2007, koll2009, koll2010, brown2014}.
On the basis of the relatively short lifetimes of giants compared
to these main sequence stars, the likelihood of observing an evolved
star in this region is even lower.

HRSs may arise from binary supernova ejection
\citep[e.g.][]{pov1967, leon1991, wang2009, tauris2015} or dynamical
ejections, boosted by Galactic rotation \citep[e.g.,][]{blaauw1961,
  dedonder1997, port2000}. Other mechanisms, such as tidal shredding of
dwarf galaxies by the Milky Way \citep{abadi2009, piffl2014} are
other possibilities.  The expected number of HRSs among early-type stars
is nonetheless somewhat lower than for HVSs \citep[e.g.,][]{perets2009, 
brown2015a}. Still, the detection of a single high-speed star with a 
Galactic disk origin, if confirmed, is likely not a strong challenge 
to theoretical predictions.

We caution that these inferences about the number counts are only
preliminary, order-of-magnitude estimates.  The \dr2\ archive is not
uniform on the sky with respect to the selection criteria we use in
deriving the sample with \qual\ astrometry.

Our reliance on the errors in parallax and proper motion to make
robust estimates for $\pub$ assumes that the noise in not sensitive to
motions not included in model fits to \gaia\ astrometric data. For
example, the \gaia\ DR2 astrometric solution does not model binary
motion \citep{lindegren2018}. Stars in binaries with orbital periods of
1--3~yr have semimajor axes of several tenths of a mas at distances of
roughly 10~kpc; this unmodeled motion could inflate the parallax,
proper motion, and associated errors. Future \gaia\ releases will
include data that cover a typical orbital period for these binaries
and a model that solves for binary motion. This analysis should
clarify the relationship between the error in the speed and the speed
for the highest velocity stars.

\section{CONCLUSION}\label{sec:conclusion}

We analyze a sample of approximately 1.5~M stars with measured radial
velocity and 5-$\sigma$ parallaxes from \gaia\ DR2 using a fast and
accurate Quasi-Monte Carlo algorithm. The code incorporates Bayesian
distance estimation and accommodates correlated erros in \dr2\ basic
source parameters.  All of the stars lie within about 15~kpc of both
the Sun and the Galactic Center.  Using their total space motion in
the Galactic rest-frame, we identify the most promising HVS and HRS
candidates. Considering only the stars' radial velocity or proper
motion, we conclude that the Galactic rest-frame radial velocity
provides a poor measure of total space motion for the fastest
stars. However, the tangential velocity alone is sufficient to
identify unbound star candidates within $\sim$15~kpc of the Sun.

We determine Galactocentric locations and speeds, along with
uncertainties, to find the probability that each source in our sample
is unbound to the Galaxy. This probability, $\pub$, depends on the
choice of Galactic potential \citep[we use the model
  in][]{kenyon2018}, the quality of the astrometric data, and the
method of distance estimation from parallax. To reduce the impact of
prior assumptions about source location on heliocentric distance
estimation, we work with sources that have relative parallax errors of
20\%\ or less. An analysis with a heliocentric distance prior based on
the bulk of \gaia\ stars gives similar results to an analysis where
all parallaxes in the error distribution out to 5-$\sigma$ give
physically plausible distances. Other assumptions, including a
constant distribution of sources in space, admit more possibilities.
We are encouraged that all methods, even the more restrictive ones,
yield the same set of stars that have a high probability of being
unbound.

However, even when a star has $\pub$ near unity, it is only one of
over 1.5 M stars with \qual\ astrometric and radial velocity data. For
stars with large measurement errors, we expect to find statistical
outliers drawn from the enormous bound population.  Thus, we introduce
an analysis to address quantitatively whether a star is truly unbound
or whether its observed kinematics are consistent with a bound
statistical outlier of a large sample. This analysis suggests that
most high speed stars in the 6-D sample are bound outliers.

Other features of the highest-speed stars support the case against 
unbound orbits. They have large errors in Galactocentric speed and
are probably late-type giants with lifetimes rather short compared to 
the time scale for unbound stars to escape the Galaxy ($\sim 100$~Myr).  
While there may be some physical explanation for the coincidence in 
timing, the idea that these stars are outliers due to the large velocity
errors is compelling.  We suspect that many of the objects identified 
by \citet{marchetti2018b} and \citet{hattori2018b}, also predominantly 
late-type giants, are bound outliers as well.

There is at least one promising object in our high-speed sample,
(\gid{5932173855446728064}), first identified by
\citet{marchetti2018b}, with the orbital elements of a star that is
unbound to the Galaxy at a high level of confidence
(\S\ref{subsec:really}). With colors (albeit reddened) that suggest an
A-type main sequence star, and an orbit that runs close to the
Galactic plane, this object is a hyper-runaway star candidate
\citep{marchetti2018b}. However, a \dr2\ error flag is set, so we
emphasize the need for observational confirmation of the source's
orbital parameters.

Twenty four other high-speed sources have trajectories and colors
consistent with late-type giants that make them improbable HVS or HRS
candidates. Our analysis of the likelihood that these objects are
unbound suggests these stars are statistical outliers of the Milky
Way's bound population.  Nonetheless, these stars are excellent
candidates for programs to obtain high quality ground-based
spectra. One of these stars, \gid{1383279090527227264}, stands out,
with the lowest probability that it is just an outlier. This object
and another star in this group (\gid{6492391900301222656}) have orbits
that passed near the LMC. Subsequent \gaia\ data releases with
improved astrometry will allow refined orbit calculations and
inferences about the origin of these high-speed stars.

Whether bound outliers or unbound stars, some of our highest-speed
stars probably have a Galactic disk origin. A significant majority
show angular momentum aligned with the Galaxy's disk
(Fig.~\ref{fig:gcvelgc}, lower right panel).  Most of this majority
are also on trajectories that are outbound from the Galactic Center.
An analysis of the type introduced here, to determine whether a source
is actually an unbound star or an outlier, may be adapted to constrain
the mass of the Milky Way inside 10-20 kpc as in \citet{gnedin2005}.

Motivated by our confirmation that proper motion alone can efficiently
select nearby high-speed stars (\S\ref{sec:getvel}; see also
\citealt{kenyon2018}), we identify new candidates selected from 5-D
\gaia\ data. Even without radial velocities, 19 stars have unbound
probabilities of 95\%\ or more, with inferred speeds between about
600~kms\ and 900~\kms. Their colors and magnitudes suggest that this
sample includes both main sequence stars as well as evolved giants.
Better astrometry and radial velocity measurements will help us learn
if these intriguing objects are among the fastest moving stars in the
Galaxy.

\acknowledgements

We thank Aaron Meisner, Anil Seth, Gail Zasowski and Zheng Zheng for
discussions about interpreting the high-velocity population of stars,
and Dustin Lang for providing access to formatted archive data. We
also thank an anonymous referee for providing comments and
suggestions that led to significant improvements in the
manuscript. This work has made use of data from the European Space
Agency (ESA) mission {\it Gaia}
(\url{https://www.cosmos.esa.int/gaia}), processed by the {\it Gaia}
Data Processing and Analysis Consortium (DPAC,
\url{https://www.cosmos.esa.int/web/gaia/dpac/consortium}). Funding
for the DPAC has been provided by national institutions, in particular
the institutions participating in the {\it Gaia} Multilateral
Agreement.  BCB and SK are grateful for generous allocations of
supercomputing time on NASA's ``discover'' cluster and the DOE's
``edison'' cluster.

\bibliographystyle{apj}
\bibliography{hvs}

\end{document}